\documentclass[preprint,prd,amsmath,amssymb,nofootinbib,tightenlines,floatfix]{revtex4}
\usepackage[dvips]{graphics}
\usepackage{epsfig}
\usepackage{latexsym}
\newcommand{\be}{\begin{equation}}
\newcommand{\ee}{\end{equation}}
\newcommand{\bea}{\begin{eqnarray}}
\newcommand{\eea}{\end{eqnarray}}
\newcommand{\lsim}{\buildrel < \over {_\sim}}

\begin{document}
\preprint{Caltech MAP-332}
\preprint{CALT-68-2648}

\title{\Large  Fermionic Effective Operators and Higgs Production at a Linear Collider}

\author{Jennifer Kile\footnote{
	Electronic address: jenkile@theory.caltech.edu}
}
\affiliation{California Institute of Technology, Pasadena, CA 91125}

\author{Michael J. Ramsey-Musolf\footnote{
	Electronic address: mjrm@caltech.edu}}
\affiliation{California Institute of Technology, Pasadena, CA 91125\\ and University of Wisconsin-Madison, Madison, WI 53706}

\begin{abstract}
We study the possible contributions of dimension six operators containing fermion fields to Higgs production at a 500 GeV or 1 TeV  $e^+e^-$ linear collider. We show that -- depending on the production mechanism -- the effects of such operators can be kinematically enhanced relative to Standard Model (SM) contributions. We determine constraints on the operator coefficients implied by existing precision electroweak measurements and the scale of neutrino mass. We find that even in the presence of such constraints, substantial deviations from SM Higgs production cross-sections are possible. We compare the effects of fermionic operators with those associated with purely bosonic operators that have been previously discussed in the literature.

\end{abstract}

\maketitle

\section{Introduction}

Uncovering the mechanism of electroweak symmetry-breaking (EWSB) will be a central goal of future experiments at the Large Hadron Collider (LHC) and the planned International Linear Collider (ILC) \cite{Heinemeyer:2005gs}. Although no direct evidence for the Standard Model Higgs boson exists and it is possible -- as in many models of EWSB -- that there exist additional scalar degrees of freedom, precision electroweak data favors at least one light scalar particle with properties akin to those of the SM Higgs boson. If it is discovered at the LHC, then measuring its properties will be an important part of the LHC and ILC program. If only a single Higgs scalar ($H$) is seen at the LHC, it is quite possible that its interactions will differ from those of the SM Higgs due to heavier degrees of freedom that are not directly accessible at the next generation of colliders. In this case, deviations of Higgs boson properties from SM expectations could provide indirect clues about the nature of physics above the TeV scale. This possibility has recently been analyzed in a model-independent way by the authors of Ref.~\cite{Barger:2003rs}, who considered  the prospective effects of dimension ($n$) six, purely (scalar) bosonic operators on $H$ production at the ILC, and in Ref.~\cite{Manohar:2006gz}, where the the potential impact of $n=6$ bosonic operators on $H$ production at the LHC were analyzed. In both cases, substantial deviations from SM expectations appear to be possible.  For recent related work, see \cite{Grinstein:2007iv}.

Here, we consider the possible impact  of $n=6$ operators containing fermions on Higgs production at a 500 GeV or 1 TeV linear collider, following the spirit of Refs. \cite{Barger:2003rs,Manohar:2006gz}. Such operators can be generated when heavy degrees of freedom, associated with a scale $\Lambda$ lying well above the EWSB scale (given by the Higgs vacuum expectation value, $v\approx 246$ GeV), are integrated out of the larger theory in which the SM is ultimately embedded. In this case, physics at low scales is described by an effective Lagrangian
\be
\label{eq:leff}
{\cal L}_{\rm eff} = \sum_{n\geq 4,\,  j} \, \frac{C_n^j}{\Lambda^{n-4}}\, {\cal O}_{n,j}\ \ \ ,
\ee
where the ${\cal O}_{n,j}$ are operators built entirely from SM fields (and possibly right-handed neutrino fields) and where the index \lq\lq $j$" runs over all independent operators of a given dimension. The operators with $n=4$ are just those of the SM (including a Dirac neutrino mass term), while the coefficients $C_n^j$ of the higher dimension operators are determined by the details of physics above the scale $\Lambda$. The effective theory described by Eq.~(\ref{eq:leff}) will be valid so long as $\Lambda >> \sqrt{s}$. 

One may analyze the possible effects of $n>4$ operators by making rather gentle assumptions about the magnitude of the operator coefficients. In the case of the $n=6$ operators of interest here, we find it useful to consider the ratio of the $C_6^j/\Lambda^2$ to the Fermi constant, $G_F=1/\sqrt{2}v^2$, that characterizes the strength of $n=6$ effective operators in the SM. Assuming that the $n=6$ operators arise from one-loop amplitudes containing particles of mass $\Lambda$, one would expect $|C_6^j/G_F\Lambda^2|\lesssim v^2/16\pi^2\Lambda^2$ or $|C_6^j v^2/\Lambda^2|\lesssim 10^{-2}$ for $v\sim\Lambda$. Taking $|C_6^j v^2/\Lambda^2|\sim 10^{-2}$, thus, gives a conservative benchmark for the magnitude of the operator coefficients\footnote{Since our effective theory is valid only when $\Lambda>>\sqrt{s} > v$, one would expect it to be applicable only when the $|C_6^j v^2/\Lambda^2|$ are much smaller than $10^{-2}$ unless the $C_6^j$ are not loop suppressed.}. In analyzing the general features  $n=6$ operator contributions to  Higgs production in $e^+e^-$ annihilation, we will generally adopt this benchmark, bearing in mind that if the new physics involves strong dynamics, the $C_6^j$ could be considerably larger\footnote{This possibility was considered more broadly in Ref.~\cite{Barger:2003rs}. See also the discussion in Ref. \cite{Black:2002wh}}. Doing so will allow us to determine which operators  may have the largest possible effects. 

After identifying the potentially most significant operators, we derive constraints on the $C_6^j v^2/\Lambda^2$ from electroweak precision observables (EWPO) and other considerations. It is well known that EWPO imply stringent bounds on operators that interfere  with the SM amplitudes for $e^+e^-\to f{\bar f}$,  and these bounds correspond to $\Lambda\gtrsim 10$ TeV or more for $C_6^j=1$ \cite{Barbieri:1999tm,Han:2004az}. Below, we update the limits obtained  in Refs.~\cite{Barbieri:1999tm,Han:2004az} on the operators with the largest prospective effects on Higgs production in $e^+e^-$ annihilation. However, operators that contain right-handed neutrino fields do not interfere with the SM amplitudes for $e^+e^-\to f{\bar f}$, and their coefficients are not all constrained by EWPO. For such operators, we turn to other considerations, such as low-energy studies of weak decays and neutrino mass \lq\lq naturalness" considerations. 

From our study of the $n=6$ operators containing both scalar and fermion fields, we arrive the following highlights: 
\begin{itemize}

\item[(i)] In contrast to the situation with purely bosonic $n=6$ operators, we show that the effects of $n=6$ operators containing fermions are generally required to be smaller, due in large part to existing precision electroweak data that agrees with SM predictions and that constrains many of the relevant operators \cite{Barbieri:1999tm,Han:2004az}. As noted above, the latter constraints are particularly strong on operators that interfere with SM amplitudes for $e^+e^-\to Z^0\to f{\bar f}$. However, we find  that substantial deviations from SM Higgs production cross-sections are possible in some cases. In particular, $n=6$ operators that contribute to the $e^+e^-\to HZ^0$ channel can generate large corrections to the SM Higgsstrahlung (HZ) cross-section at the energies considered here. The HZ cross-section can be separated from the gauge boson fusion process through appropriate choice of final states or study of the missing mass spectrum in $e^+e^- \to H \nu_e{\bar\nu}_e$. Thus, a dedicated study of HZ would provide the most sensitive probe of operators considered here. 

\item[(ii)] Although operators containing right-handed neutrino fields have not been emphasized in earlier effective operator studies of collider physics \cite{Barbieri:1999tm,Han:2004az}, the observation of neutrino oscillations and the implication of non-vanishing neutrino mass motivate us to include RH neutrinos\footnote{In doing so, we consider only Dirac neutrinos, deferring the case of Majorana neutrinos to a future study}.  Direct experimental limits on operators containing RH neutrino fields leave room for appreciable effects in Higgs production in the missing energy (${\not\!\! E}$) channel, $e^+e^-\to H +\nu{\bar \nu}$. It is possible, however, to argue for more stringent limits on these effects by invoking neutrino mass \lq\lq naturalness" considerations\cite{Bell:2005kz,Erwin:2006uc}. Below, we argue that if the only particles lighter than the SM Higgs boson are other SM particles, then the observation of large deviations from SM expectations for Higgs production with missing energy without corresponding deviations in the $H q \bar{q}$ and $H \ell \bar{\ell}$ channels would imply fine tuning in order to be consistent with the small scale of neutrino mass. 

\item[(iii)] With the possible exception of operators which would give magnetic moments to the quarks, operators containing both Higgs and quark fields, which contribute directly only to the $e^+e^-\to H {\bar q} q$ channel,  yield small contributions since their contributions are kinematically suppressed relative to SM HZ for the energies of interest here and since their operator coefficients are strongly constrained by $Z^0$ pole precision observables (except for top quarks).   While we do not directly constrain the coefficients of the quark magnetic moment operators, we find for reasonable values of these coefficients that their contributions to $e^+e^-\to H {\bar q} q$ would also be small. 

\item[(iv)] The possible effects of $n=6$ bosonic-fermionic operators are quite distinctive from those  associated with purely bosonic operators. Effects of the latter are rather generic to a variety of Higgs production channels in $e^+e^-$ annihilation, as they enter primarily through modifications of the Higgs self-couplings and Higgs coupling to gauge bosons ~\cite{Barger:2003rs} and do not change the topology or analytic properties of the Higgs production amplitudes. Moreover, these modified couplings can enter strongly in both the HZ and gauge boson fusion cross-sections and can, in principle, substantially modify the $e^+e^-\to H {\bar q} q$, $H+{\not\!\! E}$, and $H{\ell}^+{\ell}^-$ channels. In contrast, the impact of the $n=6$ operators considered here is quite channel specific, with the largest effects arising in processes dominated by SM HZ. Moreover, the analytic structure and kinematic dependence of the amplitudes generated by the $n=6$ Higgs-fermion operators is distinct from that of the SM HZ and gauge boson fusion amplitudes, a feature not associated with the purely scalar operators. Thus, a comprehensive program of Higgs production studies would provide an interesting way to disentangle the possible effects of purely bosonic and Higgs-fermion operators in Higgs production at a linear collider. 

\end{itemize}

In the remainder of the paper, we provide details of the analysis leading to these observations. In Section \ref{sec:smhiggsprod} we briefly review Higgs production in the SM. While the latter is well-known, we include a short discussion here to provide a backdrop for discussion of possible deviations from SM expectations, as the impact of the operators we consider depends strongly on both the production mechanism and energy as well as on the mass of the $H$.  Section \ref{sec:basis} contains a discussion of the $n=6$ operator basis. The heart of our study lies in Sections \ref{sec:newhiggs} and \ref{sec:oplimits} that contain, respectively, an analysis of prospective deviations from SM Higgs production due to the operators of Section \ref{sec:basis} and an evaluation of bounds on the corresponding operator coefficients obtained from various phenomenological considerations. In arriving at the latter, we follow a somewhat different procedure than used by the authors of Ref.~\cite{Barbieri:1999tm}, though the numerical differences are small. Section \ref{sec:conclusions} contains a discussion of our results and their implications.

Before proceeding, we make a few additional comments about our analysis. 
\begin{itemize}
\item[(a)] For simplicity we have considered the case of a linear collider with unpolarized beams, although the ILC will likely have one or both beams partially polarized (see Ref.~\cite{Moortgat-Pick:2005cw} and references therein).  
\item [(b)] We do not discuss changes in the Higgs production cross-section caused solely by modifications of the fermion-gauge boson  vertices in the SM Higgs production amplitudes. Effects of this type do not entail any change in the analytic structure or kinematic-dependence of the SM amplitudes, and the constraints implied by precision electroweak data and neutrino mass preclude the introduction of any significant deviations from SM Higgs production cross-sections due to changes in these couplings. 
\item[(c)] In principle, one should also consider modifications of the SM Higgs-gauge boson couplings due to contributions from $n=6$ fermionic operators to the  $\mu$-decay amplitude. The $HWW$ coupling depends on both the SU(2)$_L$ gauge coupling, $g_2$, and $M_W$, while the $HZZ$ coupling depends on $g_2$, $M_Z$, and $\cos\theta_W$, where $\theta_W$ is the weak mixing angle. The $W$ boson mass, weak mixing angle, and $g_2$ are derived quantities that depend on  the Fermi constant obtained from muon decay, corrected for $\mu$-decay dependent radiative corrections and  possible new physics contributions to the muon decay amplitude. Thus, any $n=6$ operators that contribute to the $\mu$-decay amplitude will affect the $HWW$ and $HZZ$ couplings. In practice, the constraints implied by precision electroweak data  are too strong to allow for observable effects in Higgs production cross-sections due to changes in the Higgs-gauge boson couplings generated by $n=6$ fermionic operator contributions to $\mu$-decay.  
\item[(d)] We concentrate on single Higgs production for simplicity, though the extension to $HH$ production is straightforward. 
\item[(e)] In this work, we do not consider operators that contain top quark fields.  We direct the interested reader to Ref.~\cite{Han:1999xd}. 
\end{itemize}



\section{Higgs Production in the Standard Model}
\label{sec:smhiggsprod}

\begin{figure}[h]
\epsfxsize=2in
\epsfig{figure=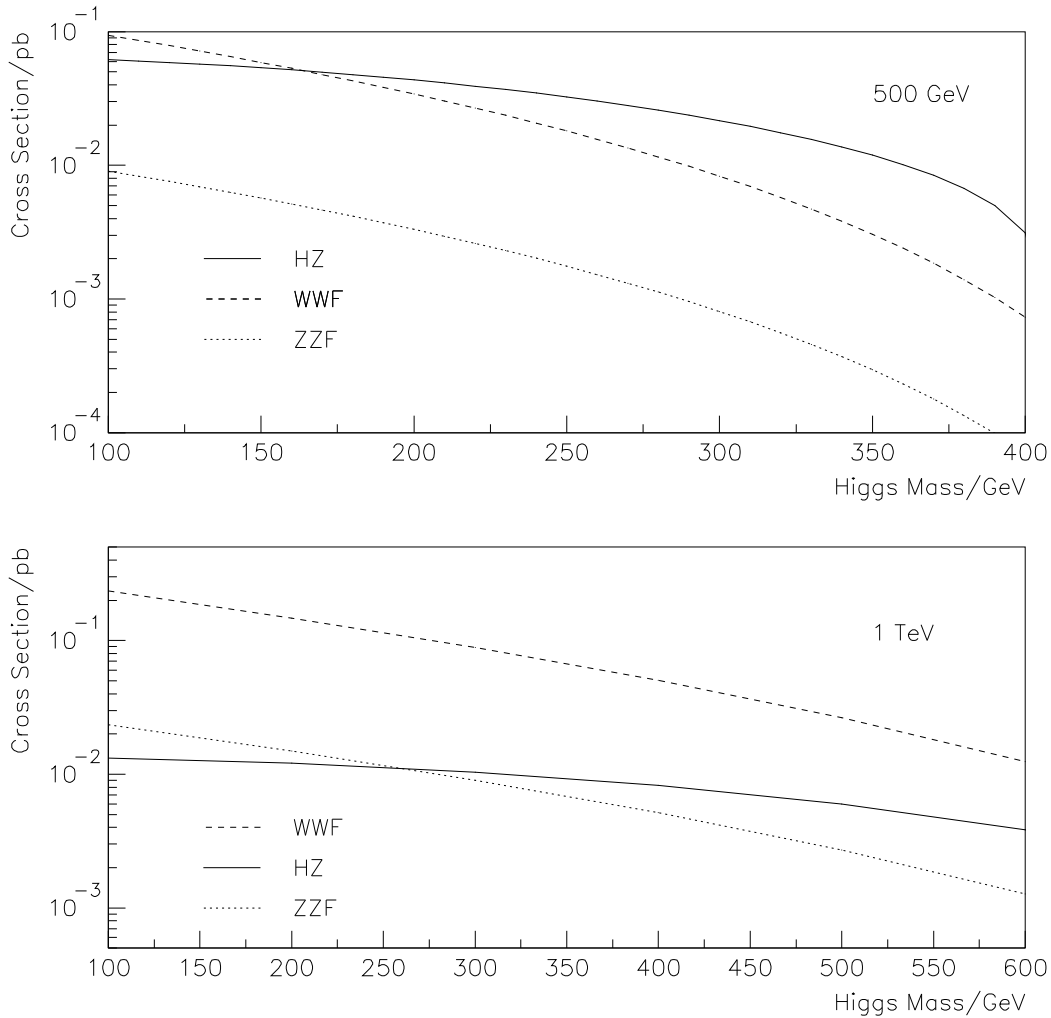,width=6.in}
\caption{SM contributions to the Higgs production cross-section.}
\label{fig:smhiggs}
\end{figure}

In the Standard Model, the Higgs boson  can be produced in $e^+ e^-$ collisions primarily by three mechanisms \cite{Gunion:1989we}.  In the Higgsstrahlung process (HZ), the $H$ is produced with an accompanying $Z^0$ boson, which then decays to a fermion-antifermion pair.  In the WW-fusion (WWF) and ZZ-fusion (ZZF) processes, the $H$ is produced with an accompanying $\nu_e \bar{\nu}_e$ and $e^+ e^-$ pair, respectively.  The cross-sections for these three processes are shown in Fig.~\ref{fig:smhiggs} for $\sqrt{s}=500$ GeV and $1$ TeV for a range of Higgs masses.  At  $\sqrt{s}=1$ TeV, the WW-fusion diagram dominates, while at $\sqrt{s}=500$ GeV, WW-fusion and Higgsstrahlung can be comparable. At lower energies (not shown here), Higgsstrahlung dominates.  The ZZ-fusion cross-section is smaller than WWF cross-section by about an order of magnitude at all energies.  Thus, for $\sqrt{s}=1$ TeV, the Higgs is primarily produced in conjunction with missing energy.  At lower $\sqrt{s}$ where HZ is important, however, one must consider final states corresponding to all possible $Z$ decay products:  $q \bar{q}$ ($70\%$), missing energy ($20\%$), and charged leptons $\ell^+ \ell^-$ ($10\%$).

In general, consideration of specific final state topologies associated with Higgs production and decay as well as $Z^0$-decay can be used to select the production mechanism. For 114 GeV $\leq m_H \lesssim 130$ GeV, the Standard Model Higgs decays primarily to $b \bar{b}$; for higher Higgs masses, the main decay channel is $W^+ W^-$.  Thus, a final state with two $b$-jets and missing energy would arise either from WWF (high $\sqrt{s}$), HZ (low $\sqrt{s}$ with $Z^0\to\nu{\bar \nu}$ and $H\to b{\bar b}$), or a combination (intermediate $\sqrt{s}$), and the corresponding event topologies at a linear collider have been studied \cite{Desch:2001at} for light values of $m_H$. The analysis of Ref.~\cite{Desch:2001at} concluded that obtaining measurement of $\sigma_{WWF}$ with $\sim 10\%$ precision or better would be feasible at a 500 GeV linear collider. 

When $H$ production is accompanied by a charged lepton-antilepton pair ($e^+ e^-$ or $\mu^+ \mu^-$ in the case of HZ and $e^+ e^-$ in the case of ZZF), the Higgs production cross-section and mass can be measured independently of its decay channel (including non-SM decays) \cite{Garcia-Abia:1999kv}.  The mass can be reconstructed from the recoil mass of the $\ell^+ \ell^-$ system.  The study of Ref.~\cite{Garcia-Abia:1999kv} considered the HZ process at $\sqrt{s}=350$ and $500$ GeV for 120 GeV $\leq m_H\leq$ 160 GeV and found that a measurement of the combined $He^+e^-$ and $H\mu^+\mu^-$ HZ cross-section with $\sim 3\%$ precision could be achieved. 
Additionally, studies have also been performed for the case of HZ where $Z\rightarrow q \bar{q}$ \cite{Garcia-Abia:2005mt,Meyer:2004ha}. In what follows, we assume that each of these event topologies can be identified experimentally, and we study the corresponding impact of $n=6$ operators assuming only SM decays of the $H$. We show that for some operators, deviations from the SM Higgs production cross-sections could be larger than the experimental error \lq\lq benchmarks" indicated above. 
 

\section{Operator Basis}
\label{sec:basis}

The basis of $n=6$ operators containing the Standard Model fields has been enumerated in previous works \cite{Leung:1984ni,Buchmuller:1985jz,Barbieri:1999tm,Han:2004az,Manohar:2006gz,Bell:2005kz,Erwin:2006uc}.  Here, we include only those containing 1) the SM Higgs doublet $\phi$ with hypercharge $Y=1$ and 2) SM fermion and/or RH neutrino fields. It is useful to distinguish three classes of such operators:  (A)  mass operators; (B) operators containing only fields that transform non-trivially under SM gauge symmetries ({\em i.e.}, do not contain $\nu_R$ fields): and (C) operators containing right-handed neutrinos that are not mass operators.\\

\vskip 0.1in

\noindent {\em Class A}.  We begin with the mass operators , of which there are two:
\bea
{\cal O}^{\ell}_{M,\, AB} &\equiv & (\bar{L}^A \phi \ell_{R}^B)(\phi^{+}\phi) \nonumber + {\rm h.c.}\\
{\cal O}^{\nu}_{M,\, AB} &\equiv & (\bar{L}^A\widetilde{\phi}\nu_{R}^B)(\phi^{+}\phi) + {\rm h.c.}\ \ \ , \nonumber
\eea
where $L^A$ and $\ell ^A$ are left-handed lepton doublet and singlet fields, respectively, and $A$, $B$ are generation indices. (Mass operators for quark fields are analogous.)  Operators containing a contracted pair of Pauli matrices, such as ${\bar L}\tau^a \phi \ell_R (\phi^\dag \tau^a \phi)$ can be related to the two operators above via a Fierz transformation.
\begin{figure}[h]
\epsfxsize=2in
\epsfig{figure=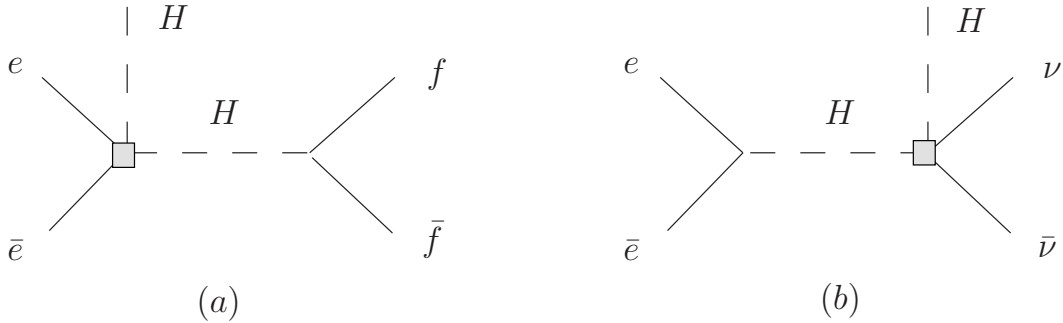,width=6.in}
\caption{Contribution of Class A operators $(a)$ ${\cal O}^{\ell}_{M,\, AB}$ and $(b)$ ${\cal O}^{\nu}_{M,\, AB}$ to Higgs production.}
\label{fig:massop1and2}
\end{figure}
The ${\cal O}^{\ell}_{M,\, AB}$ and ${\cal O}^{\nu}_{M,\, AB}$ can contribute to Higgs production via the diagrams shown in Fig.~\ref{fig:massop1and2}.  In the absence of fine-tuning with the $n=4$ Standard Model mass operators, their coefficients $C_M^{\ell}$ and $C_M^{\nu}$ are tightly constrained by the $\ell$ and $\nu$ mass, respectively:
\bea
\frac{\left| C_{M,ee}^{\ell} \right|}{\Lambda^2} &\lesssim & \frac{ 2 \sqrt{2} m_e}{v^2}\nonumber\\
\frac{\left| C_{M,AB}^{\nu} \right|}{\Lambda^2} &\lesssim & \frac{ 2 \sqrt{2}m_{\nu,AB}}{v^2}\ \ \ ,\nonumber
\eea
where $m_{\nu,AB}$ is an element of the neutrino mass matrix before diagonalization. 
In addition to this (large) suppression, the interference of these diagrams with the SM Higgs production diagrams is additionally mass-suppressed due to the fermion chiralities.  Thus, the contributions of these two operators to Higgs production are negligible, and we will not consider them further.\\

\vskip 0.1in

\noindent {\em Class B}.  These operators contain only fields that are not SM singlets ({\em i.e.}, no $\nu_R$): 

\bea
{\cal O}_{VR,AB} &\equiv& i(\bar{f}_{R}^A\gamma^{\mu}f_{R}^B)(\phi^{+}D_{\mu}\phi) + {\rm h.c.} \nonumber \\
{\cal O}_{VL,AB} &\equiv& i(\bar{F}^A\gamma^{\mu}F^B)(\phi^{+}D_{\mu}\phi) \nonumber  + {\rm h.c.}\\
{\cal O}_{VL\tau,AB} &\equiv& i(\bar{F}^A\gamma^{\mu}\tau^{a}F^B)(\phi^{+}\tau^{a}D_{\mu}\phi) + {\rm h.c.}\nonumber \\
{\cal O}_{{\tilde V},\, AB}^q  &\equiv& i(\bar{d}_{R}^A\gamma^{\mu}u_{R}^B)(\phi^{+}D_{\mu}\widetilde{\phi}) + {\rm h.c.} \nonumber \\
{\cal O}_{W,AB}^{f} &\equiv& g_{2}(\bar{F}^A \sigma^{\mu\nu}\tau^{a}\phi)f_{R}^B W_{\mu\nu}^{a} + {\rm h.c.} \nonumber \\
{\cal O}_{B,AB}^{f} &\equiv& g_{1}(\bar{F}^A \sigma^{\mu\nu}\phi)f_{R}^B B_{\mu\nu} + {\rm h.c.}\ \ \ , \nonumber 
\eea
where $F^A$ indicates either the left-handed lepton ($L$) or quark ($Q$) doublet for generation $A$ and $f^A$ indicates the RH fields for quarks or charged leptons of generation $A$. We have included the \lq\lq $R$" subscript on the latter for clarity.  The fields $u_R^A$ and $d_R^A$ denote the up- and down-type RH quarks of generation $A$. The operator ${\cal O}_{{\tilde V},\, AB}^{q} $ does not contribute to Higgs production in $e^+e^-$ annihilation since it contains no neutral current component, so we will not discuss it further.

\vskip 0.1in

\noindent{\em Class C}. Lastly, we consider operators containing $\nu_R$ that are not mass-suppressed and that contribute only to the missing energy channel:

\bea
{\cal O}_{V\nu,\,AB} &\equiv& i(\bar{\nu}_{R}^A\gamma^{\mu}\nu_{R}^B)(\phi^{+}D_{\mu}\phi) + h.c. \nonumber \\
{\cal O}_{{\tilde V},\, AB}  &\equiv& i(\bar{\ell}_{R}^A\gamma^{\mu}\nu_{R}^B)(\phi^{+}D_{\mu}\widetilde{\phi}) + {\rm h.c.} \nonumber \\
{\cal O}_{W,\,AB} &\equiv& g_{2}(\bar{L}^A \sigma^{\mu\nu}\tau^{a}\widetilde{\phi})\nu_{R}^B W_{\mu\nu}^{a} + {\rm h.c.} \nonumber \\
{\cal O}_{B,\,AB} &\equiv& g_{1}(\bar{L}^A \sigma^{\mu\nu}\widetilde{\phi})\nu_{R}^B B_{\mu\nu} + {\rm h.c.} \nonumber 
\eea
For ${\cal O}_{{\tilde V},\, AB}$, ${\cal O}_{W,\,AB}$, and ${\cal O}_{B,\,AB}$, we follow the notation of Refs.~\cite{Bell:2005kz,Erwin:2006uc}.  Due to the presence of the $\nu_R$ field, interference of tree-level diagrams containing these operators with the Standard Model Higgs production amplitudes is suppressed by the neutrino mass.  
Hence, we do not consider these interference effects here and compute only the contributions that are quadratic in their coefficients. As a result, their contributions can be appreciable only if the corresponding $C_6^j$ are not loop suppressed.


\section{Contributions to Higgs Production}
\label{sec:newhiggs}

\subsection{General Considerations}

\begin{figure}[h]
\epsfxsize=2in
\epsfig{figure=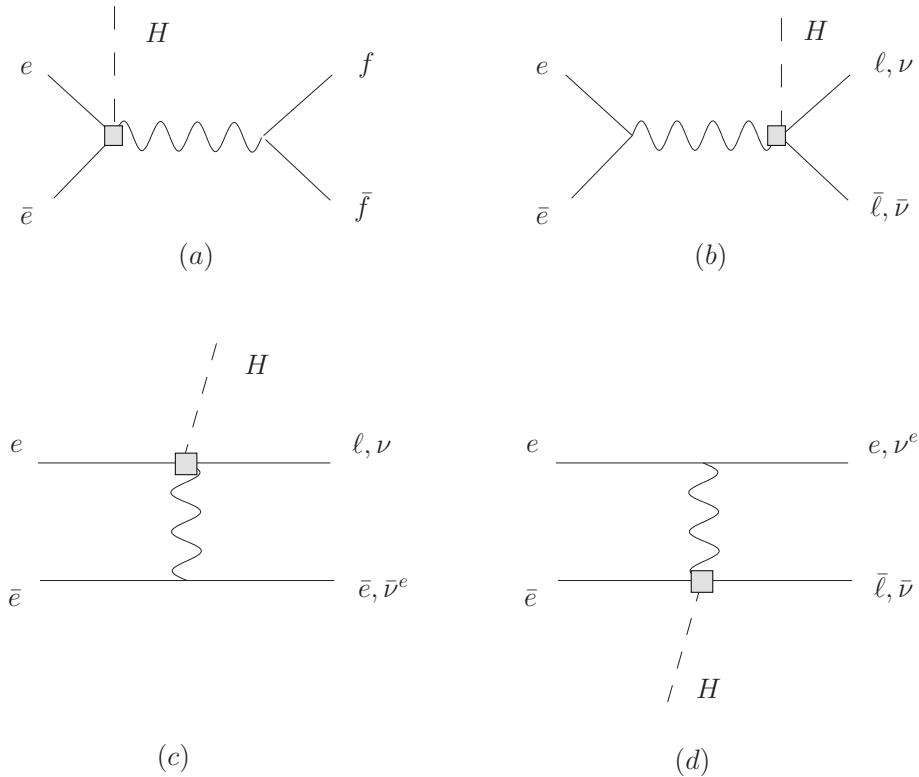,width=5.in}
\caption{Contribution of Class B operators to Higgs production.}
\label{fig:nonur}
\end{figure}

\begin{figure}[h]
\epsfxsize=2in
\epsfig{figure=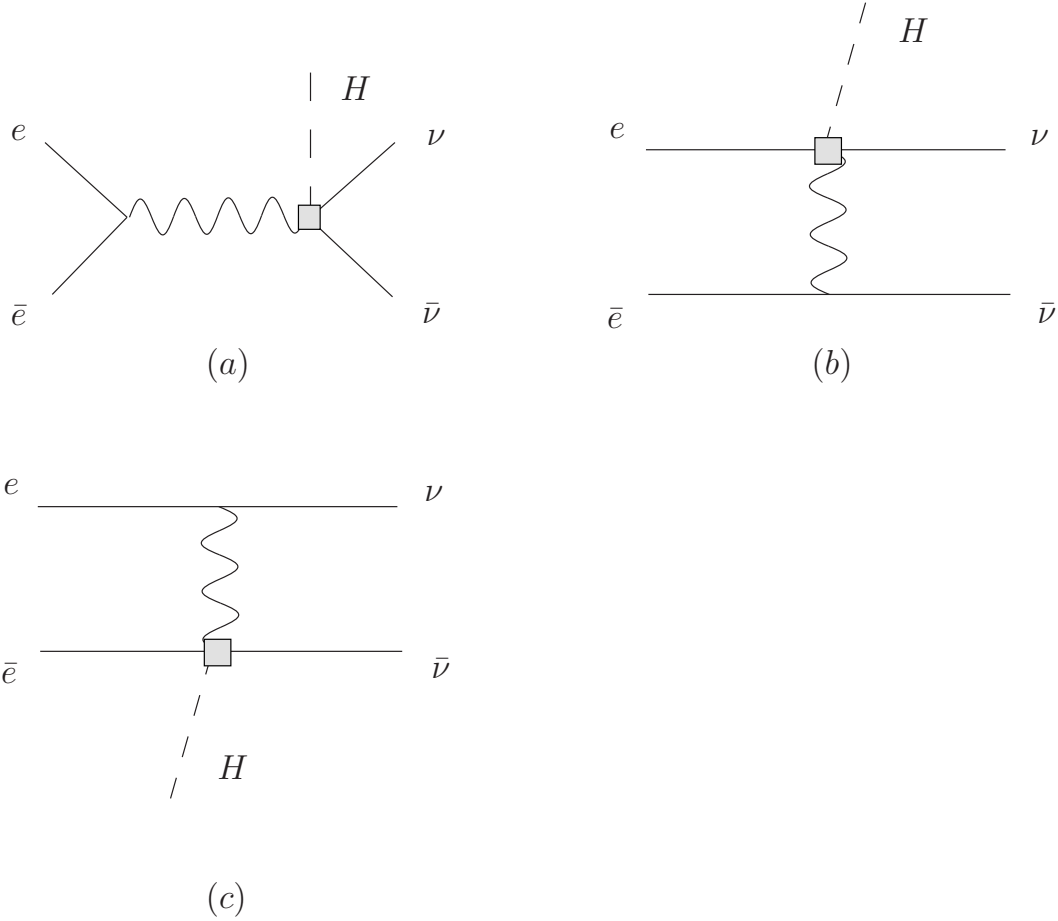,width=6.in}
\caption{Contribution of Class C operators to Higgs production.}
\label{fig:nur}
\end{figure}

Before considering in detail the corrections to various production channels, we make a few general observations regarding the operators and amplitudes that one may expect to be largest. To that end, we show in Figure \ref{fig:nonur} the $H$ production amplitudes generated by the operators of Class B and in Figure \ref{fig:nur} those generated by Class C operators. The amplitudes in Figs. \ref{fig:nonur}(a,b) and \ref{fig:nur}(a) correspond to taking the SM HZ amplitude and contracting one of the two $Z^0$ propagators to a point. In SM HZ, the initial $Z^0$ is far off shell  for the energies considered here, while the final $Z^0$ propagator is resonant. Thus, we expect the contributions associated with Figs. \ref{fig:nonur}(b) and \ref{fig:nur}(a) to be highly suppressed relative to the SM cross-section since they contain no resonating $Z^0$ propagator. In contrast, the amplitude of Fig. \ref{fig:nonur}(a)  contains a nearly on-shell $Z^0$ propagator but no off shell $Z^0$ propagator. Consequently, it can be kinematically enhanced relative to the SM HZ amplitude and 
can generate an appreciable contribution to $H$ production, even in the presence of strong constraints on the corresponding operator coefficient (see Sec. \ref{sec:oplimits}). 

The corrections generated by the amplitudes of Figs. \ref{fig:nonur}(c,d) and \ref{fig:nur}(b,c)  contribute to the  $Hl^A{\bar l}^B$ (where at least one of $A$ and $B=e$) and missing energy channels. For large $\sqrt{s}$, the $H+{\not\!\! E}$ channel is dominated by WWF wherein both $W$ bosons are off shell. Thus, the amplitudes of Figs. \ref{fig:nonur}(c,d) and \ref{fig:nur}(b,c) experience no kinematic suppression relative to the SM cross-section\footnote{This situation contrasts with that of Fig. \ref{fig:nonur}(b), which corresponds to shrinking the resonating $Z^0$ propagator in HZ to a point, thus leading to a kinematic suppression relative to the SM HZ amplitude.}.  Even in the intermediate energy regime, where WWF and HZ yield comparable contributions, the effects of Figs. \ref{fig:nonur}(c,d) and \ref{fig:nur}(b,c) can, in principle, be appreciable. We reiterate, however, that for the operators containing $\nu_R$ fields, the amplitudes of Fig. \ref{fig:nur} do not interfere appreciably with the SM amplitudes, and their contributions can only be large when the operator coefficients are not loop-suppressed.

We now turn to a detailed discussion of various operator effects. 

\subsection{Class B Operators}

Here, we discuss in detail the possible effects of operators in Class B, which contain only fields that transform non-trivially under SM symmetries.

\vskip 0.1in

\noindent{${\cal O}_{VR,AB}$}

\vskip 0.1in

The contributions from operator ${\cal O}_{VR,AB}$ depends on its flavor indices $A, B$. For $A=B=e$, ${\cal O}_{VR,ee}$ contributes to all Higgs production channels via the diagram in Fig. \ref{fig:nonur} (a) and additionally to the $H e^+ e^-$ channel via the diagrams in \ref{fig:nonur} (b-d).  In all cases, the exchanged gauge boson is a $Z^0$.    As noted above, the analytic structure of the amplitude for Fig. \ref{fig:nonur}(a) differs from that of the SM HZ amplitude only by the absence of the off-shell $Z^0$ propagator. The ratio of its interference with the SM HZ amplitude to the SM HZ cross-section is, thus, given by 
\be
\frac{\sigma_{3(a)-HZ\, {\rm int}}}{\sigma_{HZ}} =   - \frac{C v^2}{\Lambda^2}  \frac{(s-M^2_Z)}{M^2_Z} \frac{\sin^2 \theta_W}{2(\sin^4 \theta_W- \frac{1}{2} \sin^2 \theta_W + \frac{1}{8})}\ \ \ ,
\label{eq:2lrhzint}
\ee
where we have omitted the label on the operator coefficient for simplicity. 
For $C v^2/\Lambda^2 =10^{-2}$, this ratio is $\sim -0.54$ and $\sim -2.2$  for $\sqrt{s}=500$ GeV and $1$ TeV, respectively.  The effect of $\sigma_{3(a)-HZ \, {\rm int}}$ relative to $\sigma_{HZ}$ can be large for the values of $\sqrt{s}$ studied here since in the SM HZ amplitude the initial $Z^0$ is far off shell with $M_Z \ll \sqrt{s}$; thus, the SM HZ amplitude contains a kinematic suppression of roughly $\Lambda^2/s$ that does not enter the amplitude of Fig. \ref{fig:nonur}(a). 

For any of the final states of $H f\bar{f}$ with $f=\mu$, $\tau$, $\nu_\mu$, $\nu_\tau$, or $q$, Eq. (\ref{eq:2lrhzint})  gives the ratio of the contribution of ${\cal O}_{VR,ee}$ to the SM cross-section.  For the $H \nu_e \bar{\nu_e}$ final state, the SM also receives a contribution from the WWF process\footnote{Since the neutrinos in the missing energy channel are not detected, one may discuss the relative magnitudes of non-SM contributions using the neutrino flavor basis.}.  Interference between WWF -- which involves only a LH (RH) initial state electron (positron) -- and diagram \ref{fig:nonur}(a) containing ${\cal O}_{VR,ee}$ requires a Yukawa coupling on each of the initial-state fermion lines, and is thus strongly suppressed.  For the $H e^+ e^-$ production channel, we must include the interference of all of the diagrams shown in Fig. \ref{fig:nonur} with both SM HZ and ZZF.  

\begin{figure}[h]
\epsfxsize=2in
\epsfig{figure=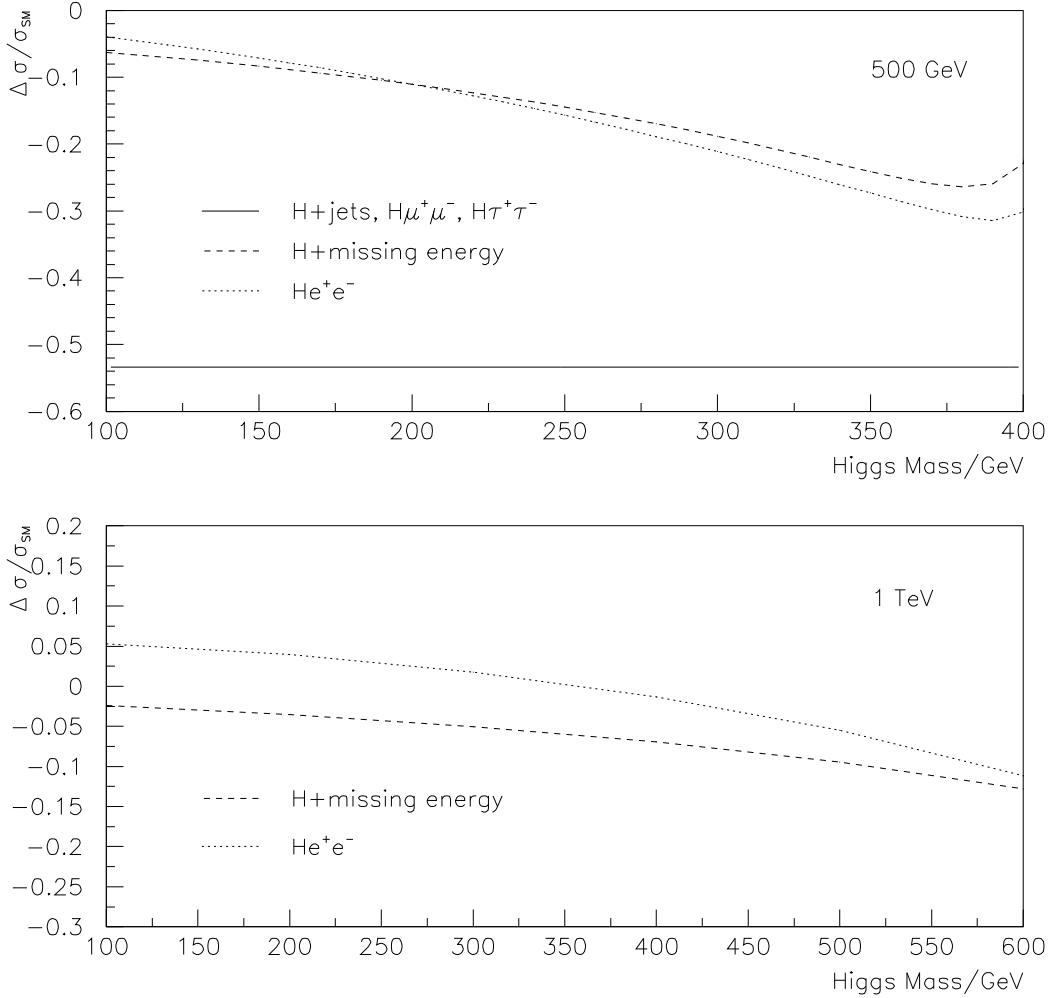,width=6.in}
\caption{Ratio of contribution of ${\cal O}_{VR,ee}$ to SM Higgs production cross-section for $(top)$ $\sqrt{s}=500$ GeV and $(bottom)$ $1$ TeV for $C_{VR,ee} v^2/{\Lambda^2}=10^{-2}$.  For $\sqrt{s}=1$ TeV, the line for the $Hq\bar{q}$, $H\mu^+\mu-$ and $H\tau^+\tau^-$ channels is not shown; it has the value of $-2.2$, independent of Higgs mass.}
\label{fig:2lree500}
\end{figure}

We have computed the contribution of ${\cal O}_{VR,ee}$ arising from interference with the SM amplitudes\footnote{ Here, we neglect the contributions that are not due to interference with the SM; we will defer discussion of the non-interference terms to Section \ref{sec:conclusions}.} to the total $H$ production cross-section using the calchep package \cite{Pukhov:1999gg,Pukhov:2004ca}. Results are shown in  Fig. \ref{fig:2lree500}, where we give the ratio $\sigma_{{\rm int}}/{\sigma_{\rm SM}}$  as a function of the Higgs mass for different final state topologies, where $\sigma_{{\rm int}}$ is the contribution to the cross-section of the interference between all of the diagrams in Fig. \ref{fig:nonur} and all of the relevant SM diagrams.  We observe that for the  $H f\bar{f}$ channels with $f=\mu$, $\tau$, $\nu_\mu$, $\nu_\tau$, or $q$, the ratio is independent of $m_H$, as implied by Eq.~(\ref{eq:2lrhzint}). In contrast, for the $He^+e^-$ and $H+{\not\!\! E}$ channels, the ratio varies with $m_H$ due to the additional contributions from the SM WWF and ZZF processes as well as other diagrams in Fig \ref{fig:nonur}. We also note that the effect of ${\cal O}_{VR,ee}$ can be large compared with the SM HZ cross-section. Thus, one could in principle discern the effects of this operator by analyzing events that cannot be produced by the WWF process, such as a dilepton pair and two $b$-jets or two $b$-jets and two other jets. In contrast, the relative effect of ${\cal O}_{VR,ee}$ on the $He^+e^-$ and $H+{\not\!\! E}$  channels is considerably smaller, due to the much larger SM ZZF and WWF  contributions in these cases. 

\begin{figure}[h]
\epsfxsize=2in
\epsfig{figure=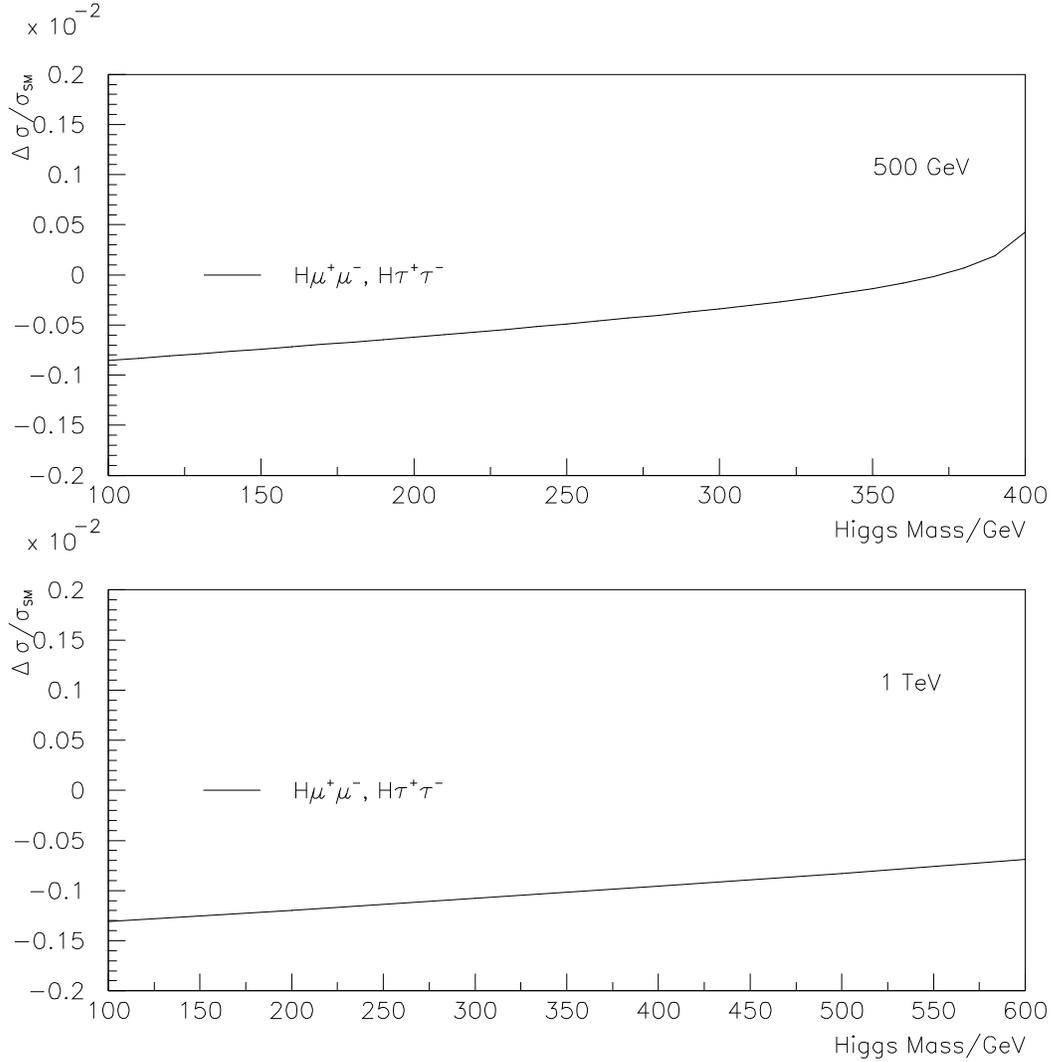,width=6.in}
\caption{Ratio of contribution of ${\cal O}_{VR,\mu \mu}$ to SM Higgs production cross-section for $(top)$ $\sqrt{s}=500$ GeV and $(bottom)$ $1$ TeV for $C_{VR,\mu \mu} v^2/\Lambda^2=10^{-2}$.  Curves for ${\cal O}_{VR,\tau \tau}$ are identical. }
\label{fig:2lrmumu500}
\end{figure}

In contrast to the situation with ${\cal O}_{VR,ee}$, the operator ${\cal O}_{VR,AA}$, $A = \mu, \tau,q$ contributes only through diagram \ref{fig:nonur}(b).  This diagram interferes only with the HZ amplitude  and contributes  only to the $H\mu^+\mu^-$, $H\tau^+\tau^-$ and $Hq \bar{q}$  channels.  The contribution of ${\cal O}_{VR,\mu\mu}$ to the $H\mu^+\mu^-$ channel -- relative to the SM cross-section -- is shown in Fig.~\ref{fig:2lrmumu500} as a function of $m_{H}$.  The results for ${\cal O}_{VR,\tau\tau}$ are identical; those for ${\cal O}_{VR,qq}$($q\not=t$) differ from Fig.~\ref{fig:2lrmumu500} only due to the difference between the $Zqq$ and $Z\ell^+\ell^-$ SM couplings. As indicated in Fig.~\ref{fig:2lrmumu500}, the contribution from ${\cal O}_{VR,\mu\mu}$ to the $H\mu^+\mu^-$ channel is $\lsim 10^{-3}$ of the SM cross-section, and we do not show the correspondingly small correction from  ${\cal O}_{VR,qq}$ to the $Hq{\bar q}$ channel. 

Comparing the contributions of ${\cal O}_{VR,ee}$ and ${\cal O}_{VR,\mu\mu}$ to the $H\mu^+ \mu^-$ channel in Figs.~\ref{fig:2lree500} and ~\ref{fig:2lrmumu500}, we can see that the effects of diagram \ref{fig:nonur}(b) are strongly suppressed relative to those of diagram \ref{fig:nonur}(a).  
As noted above, this suppression is to be expected, since in the amplitude of Fig. \ref{fig:nonur}(b)
the $Z^0$ is always off-shell ($M_Z \ll \sqrt{s}$), whereas for the values of $\sqrt{s}$ of interest here, on-shell production of both the $H$ and $Z^0$ can occur for the amplitude of Fig. \ref{fig:nonur}(a). As the same arguments will hold for ${\cal O}_{VL,AB}$ and ${\cal O}_{VL\tau}$, we will not consider the case of $A=B=\mu,\tau$ for those operators below.\\

\vskip 0.1in

\noindent{${\cal O}_{VL,ee}$}

\vskip 0.1in

\begin{figure}[h]
\epsfxsize=2in
\epsfig{figure=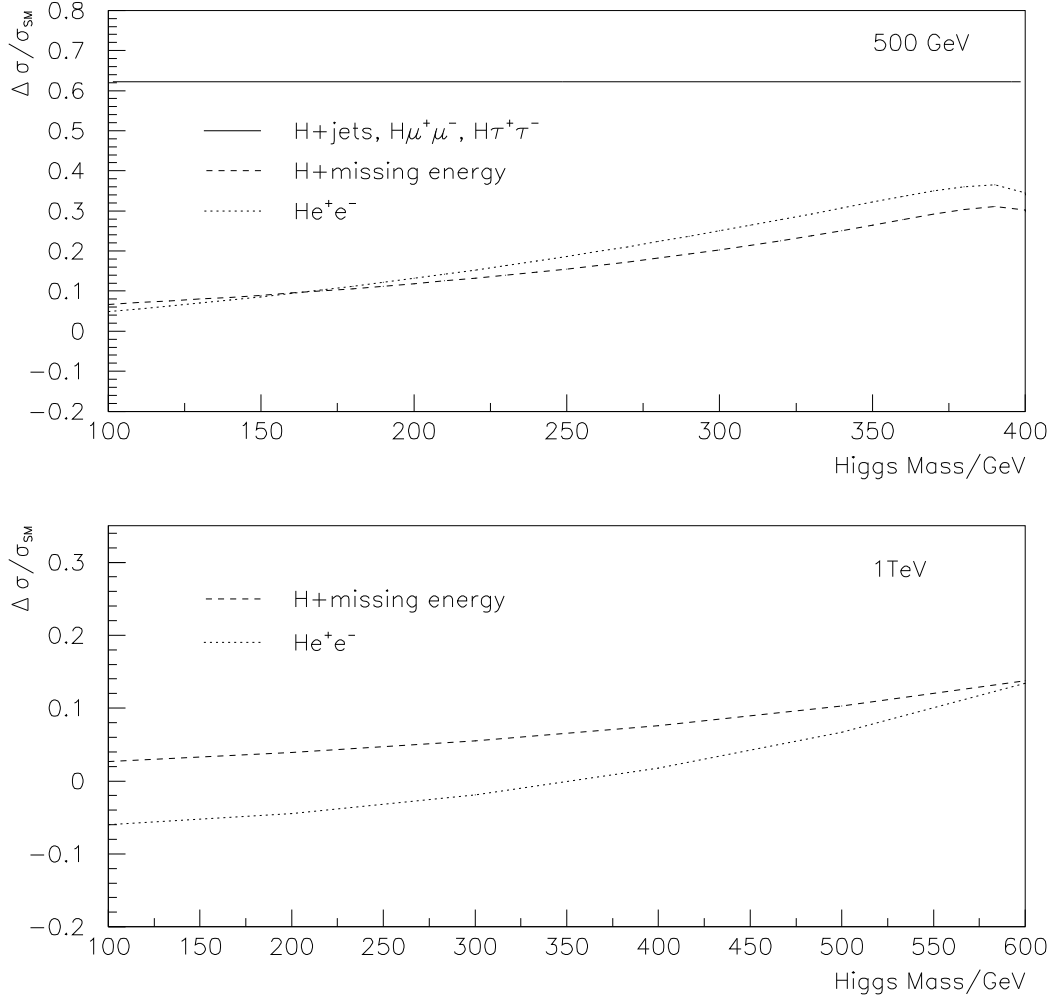,width=6.in}
\caption{Ratio of contribution of ${\cal O}_{VL,ee}$ to SM Higgs production cross-section for (top) $\sqrt{s}=500$ GeV and (bottom) $\sqrt{s}=1$ TeV for $C_{VL,ee} v^2/\Lambda^2=10^{-2}$.  For $\sqrt{s}=1$ TeV, the line for the $Hq\bar{q}$, $H\mu^+\mu-$ and $H\tau^+\tau^-$ channels is not shown; it has the value of $2.6$, independent of Higgs mass.}
\label{fig:2Lee01}
\end{figure}

As with ${\cal O}_{VR,ee}$, the operator ${\cal O}_{VL,ee}$ contributes to Higgs production via the diagrams in Fig. \ref{fig:nonur}(a-d).  In all four diagrams, the gauge boson exchanged is always a $Z^0$.  Diagram \ref{fig:nonur}(a) contributes to all channels, in analogy with ${\cal O}_{VR,ee}$ above.  This contribution of the interference of this diagram with HZ obeys
\be
\label{eq:2llhzint}
\frac{\sigma_{3(a)-HZ int}}{\sigma_{HZ}} =    \frac{Cv^2}{\Lambda^2}  \frac{(s-M^2_Z)}{M^2_Z} \frac{(\frac{1}{2}-\sin^2 \theta_W)}{2(\sin^4 \theta_W- \frac{1}{2} \sin^2 \theta_W + \frac{1}{8})}
\ee
This expression  gives the ratio of the contribution of ${\cal O}_{VL,ee}$-SM HZ interference to the SM cross-section for the final states of $Hf\bar{f}$ for $f=\mu$, $\tau$, $\nu_{\mu,\tau}$, and $q$. However, in contrast to the situation with  ${\cal O}_{VR,ee}$, the insertion of this operator diagram \ref{fig:nonur}(a) will also interfere with WWF without electron mass insertions (as well as with HZ and ZZF).   Additionally, ${\cal O}_{VL,ee}$ contributes to the $H e^+ e^-$ channel through diagrams \ref{fig:nonur}(b-d), all of which interfere with HZ and ZZF, and to the $H \nu_e \bar{\nu_e}$ through diagram \ref{fig:nonur}(b) (although this latter contribution is strongly kinematically suppressed for the reasons discussed above).  These contributions are summarized in Fig. \ref{fig:2Lee01} for $C v^2/\Lambda^2=10^{-2}$ as a function of $m_{H}$. As before, the relative effect on the $Hf\bar{f}$ cross-section is $m_H$-independent for $f=\mu$, $\tau$, $\nu_{\mu,\tau}$, and $q$, whereas for the $H e^+ e^-$ and $H+{\not\!\! E}$ channels, the relative importance decreases with $m_H$ owing to the increasing ZZF and WWF contributions.

As in the case of ${\cal O}_{VR,AA}$, the contribution from ${\cal O}_{VL,AA}$ for $A=\mu$, $\tau$, or $q$ arises only from Fig. \ref{fig:nonur}(b). Since the corresponding effects are highly suppressed, we do not discuss this case further.

\vskip 0.1in

\noindent{${\cal O}_{VL\tau,ee}$}

\vskip 0.1in

As in the previous cases, ${\cal O}_{VL\tau,ee}$ contributes to the Higgs production cross-section through all of the diagrams in Fig.~\ref{fig:nonur}.  However, unlike the operators ${\cal O}_{VR,ee}$ and ${\cal O}_{VL,ee}$, ${\cal O}_{VL\tau,ee}$ also contains a charge-changing component.  Thus, the gauge boson in diagrams \ref{fig:nonur}(c) and (d) can be either a $Z^0$ or a $W^\pm$, so the insertion of ${\cal O}_{VL\tau,ee}$ in these diagrams contributes to both the $He^+e^-$ and $H+{\not\!\! E}$ channels.

\begin{figure}[h]
\epsfxsize=2in
\epsfig{figure=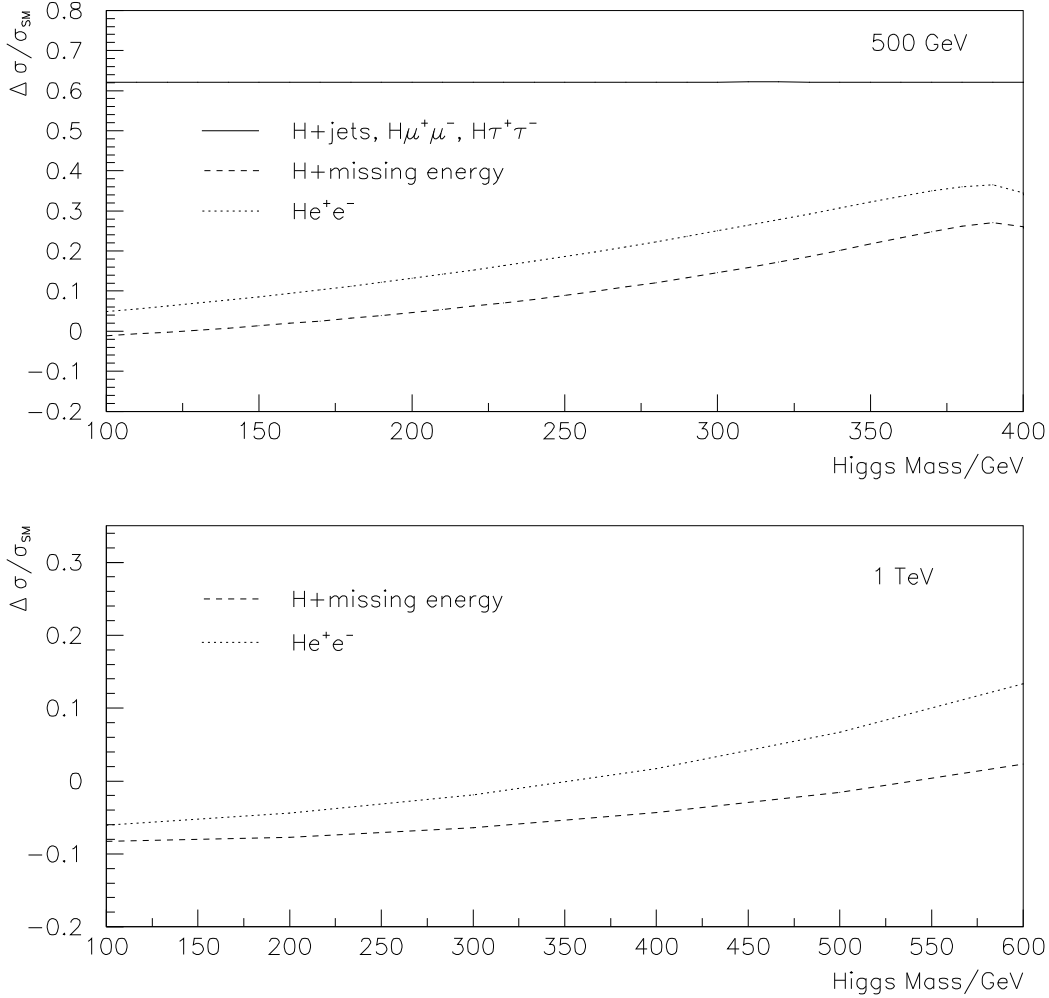,width=6.in}
\caption{Ratio of contribution of ${\cal O}_{VL\tau,ee}$ to SM Higgs production cross-section for (top) $\sqrt{s}=500$ GeV and (bottom) $1$ TeV for $C_{VL\tau,ee} v^2/\Lambda^2=10^{-2}$.  For $\sqrt{s}=1$ TeV, the line for the $Hq\bar{q}$, $H\mu^+\mu-$ and $H\tau^+\tau^-$ channels is not shown; it has the value of $2.6$, independent of Higgs mass.}
\label{fig:2Ltauee01}
\end{figure}

Inserting ${\cal O}_{VL\tau,ee}$ in diagram \ref{fig:nonur}(a) generates the same contribution
to all decay channels in the same manner as $O_{VL,ee}$, yielding the same contribution to the HZ cross-section as for ${\cal O}_{VL\tau,ee}$ (see, {\em e.g.}, Eq.~(\ref{eq:2llhzint})).
The insertion of ${\cal O}_{VL\tau,ee}$ in diagram \ref{fig:nonur}(a) also interferes with ZZF and WWF in the $H e^+ e^-$ and $H \nu_e \bar{\nu_e}$ channels, respectively. Additionally, ${\cal O}_{VL\tau,ee}$ contributes to these channels via diagrams \ref{fig:nonur}(b-d).  The contributions of ${\cal O}_{VL\tau,ee}$ to the Higgs production cross-section are are shown in Fig. \ref{fig:2Ltauee01} for $C v^2/\Lambda^2=10^{-2}$.

As in the case of ${\cal O}_{VR,AA}$, the contribution from ${\cal O}_{VL\tau,AA}$ for $A=\mu$, $\tau$, or $q$ arises only from Fig. \ref{fig:nonur}(b). Since the corresponding effects are highly suppressed, we do not discuss this case further. 

\vskip 0.1in

\noindent{${\cal O}_{W,AB}^{f}$ and ${\cal O}_{B,AB}^{f}$}

\vskip 0.1in

The operators ${\cal O}_{W}^{f}$ and ${\cal O}_{B}^{f}$ contribute to the magnetic and electric dipole moments of the charged leptons.  Stringent limits on the electric dipole moments and non-SM contributions to the magnetic moments exist for the cases $A=B=e$ and $A=B=\mu$ \cite{Yao:2006px}.  Limits on the branching fractions $\mu \rightarrow e \gamma$, $\tau \rightarrow e \gamma$, and $\tau \rightarrow \mu \gamma$ tightly constrain the cases where $A$ and $B$ are lepton fields and $A\ne B$ \cite{Yao:2006px}.  Thus, here we will only consider the possibilities $A=B=\tau$ and $A,B=q^Aq^B$.  

${\cal O}_{W,\tau\tau}^{f}$ and ${\cal O}_{B,\tau\tau}^{f}$ will contribute only to the $H\tau^+ \tau^-$ final state; production occurs only through diagram \ref{fig:nonur} (b).  Due to the derivative on the gauge boson field in each of these operators, the kinematic suppression of this diagram is not as severe as in the previous cases of ${\cal O}_{VR,AB}$, ${\cal O}_{VL,AB}$ and ${\cal O}_{VL\tau,AB}$.  

We have calculated the contributions of ${\cal O}_{W,\tau\tau}^{f}$ and ${\cal O}_{B,\tau\tau}^{s}$ to the $H\tau^+ \tau^-$ cross-section for $C^j v^2/\Lambda^2=10^{-2}$, neglecting the Yukawa-suppressed contribution to the cross-section due to the interference of diagram \ref{fig:nonur} (b) with the SM HZ process.  We find that the contribution to the cross-section is generally less than $0.1\%$ for $\sqrt{s}=500$ GeV, and less than $2\%$ for $\sqrt{s}=1$ TeV.  We also find that the interference of diagram \ref{fig:nonur} (b) with other (tiny) SM processes which contain a Higgs insertion on one of the $\tau$ lines could give comparable contributions to the $H\tau^+ \tau^-$ cross-section.

For the case where $A$ and $B$ are light quark fields ($u$, $d$, and $s$), interference with the SM diagrams can be neglected as these contributions are Yukawa-suppressed.
There is a contribution to the $Hq^A\bar{q}^B$ cross-section that is $N_C=3$ times larger than the $A=B=\tau$ noninterference cross section discussed above and is, thus, negligible .  In the case where $A=B=b$ or $c$, interference with the SM diagrams can give additional contributions with magnitude comparable to the non-intereference contributions.

Current limits \cite{Yao:2006px} on the $\tau$ magnetic moment allow values for $C_{B,\tau\tau}^{f}v^2/\Lambda^2$ and $C_{W,\tau\tau}^{f}v^2/\Lambda^2$ of order unity.  Somewhat improved limits, but still significantly weaker than $C_{B,W,\tau\tau}^{f}v^2/\Lambda^2=10^{-2}$ can be obtained from $\Gamma(Z\rightarrow \tau^+ \tau^-)$.  Similarly weak limits on the quark magnetic moment operators can be obtained from $\Gamma(Z\rightarrow q^A \bar{q}^B)$.  However, we will take $10^{-2}$ as an estimate of the upper bound for $C_{B,W}^f v^2/\Lambda^2$, as we do not expect new physics to make a contribution to the magnetic moments greater than the QED Schwinger term. Nevertheless, we do not rule out the possibility that the coefficients of these operators could be considerably larger due to strong dynamics above the scale $\Lambda$.



\subsection{Class C Operators}

All of the Class C operators  contribute only to the missing energy channel since they contain $\nu_R$ fields.  The Higgs production diagrams for these operators  are shown in Fig.~\ref{fig:nur}. For each operator, the interference of any amplitude in Fig.~\ref{fig:nur} with relevant SM amplitude is $m_\nu$-suppressed, so we do not include the interference contributions here.  The resulting corrections to the SM Higgs production cross-sections are, thus, quadratic in the operator coefficients. 

\begin{figure}[h]
\epsfxsize=2in
\epsfig{figure=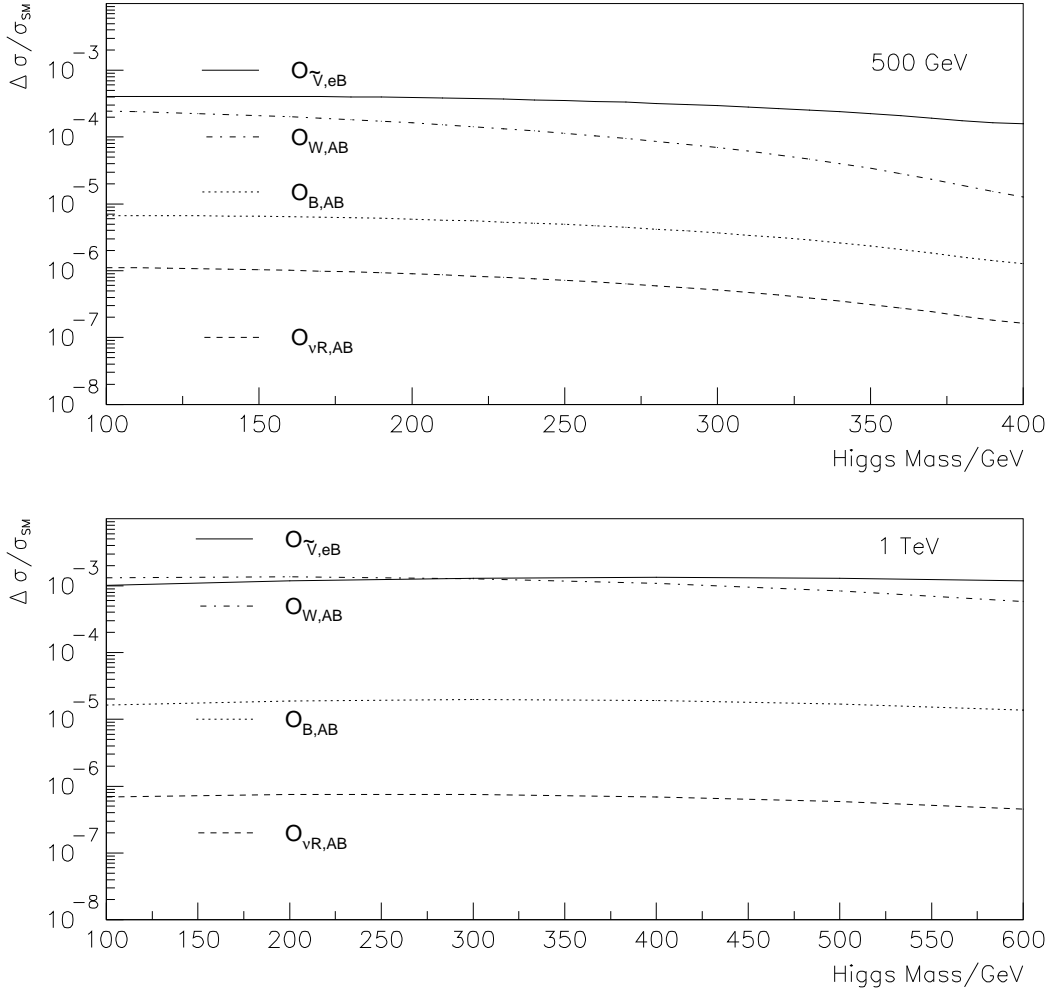,width=6.in}
\caption{Contributions of operators containing $\nu_R$ to Higgs missing energy final state for $\sqrt{s}=500$ GeV .  Results are as a fraction of total the Standard Model $H \nu \bar{\nu}$ cross-section, summed over the three flavors.  Curves are drawn for the case $C^j v^2/\Lambda^2=10^{-2}$.}
\label{fig:nur500Gev}
\end{figure}

Since the final state neutrino-antineutrino pair is not observed, we do not require their flavors to be the same. As discussed above, the contribution from diagram \ref{fig:nur}(a) in is kinematically suppressed due to the off-shell $Z^0$ boson, so we expect that only those operators contributing through diagrams \ref{fig:nur}(b) and (c) will be able to generate substantial contributions.  The comparison between the contribution from  these operators to the $H+{\not\!\! E}$ channel is given in Fig. \ref{fig:nur500Gev} for $C v^2/\Lambda^2=10^{-2}$. 

For $C v^2/\Lambda^2=10^{-2}$ as assumed above, the correction induced by the Class C operators is generally less than $10^{-3}$ of the SM cross-section.  However, if these operators are generated by strong dynamics or tree-level gauge interactions, their relative effects could be substantially larger. In this respect, the operator ${\cal O}_{\tilde{V},AB}$ is particularly interesting, as an operator of this type could arise in models with mixing between LH and RH gauge bosons. Moreover, it is not as strongly constrained by precision electroweak data as the Class B operators, since it does not interfere with the SM amplitudes that contain only LH neutrino fields. In Section \ref{sec:oplimits} we discuss the various phenomenological  and theoretical constraints on ${\cal O}_{\tilde{V},AB}$, including implied by the scale of neutrino mass and naturalness considerations. 


\vskip 0.1in

\noindent{${\cal O}_{V\nu,AB}$}

\vskip 0.1in

The operator ${\cal O}_{V\nu,AB}$ contributes to the missing energy channel only via the diagram in Fig. \ref{fig:nur}(a) where the exchanged gauge boson is a $Z^0$ and the final state contains a right-handed neutrino and a left-handed antineutrino.  Thus, the contribution of this operator is strongly kinematically suppressed,  as reflected in Fig.~\ref{fig:nur500Gev}.\\

\vskip 0.1in

\noindent{${\cal O}_{\tilde{V},AB}$}

\vskip 0.1in

The gauge boson in ${\cal O}_{\tilde{V},AB}$ is always a $W^\pm$, and this operator contributes to the missing energy channel via the diagrams in Fig. \ref{fig:nur} (b) and (c).  The final state contains one right-handed neutrino and one right-handed antineutrino, in the case of \ref{fig:nur}(b), or a left-handed neutrino and antineutrino in the case of \ref{fig:nur}(c).  As this operator contributes through diagrams (b) and (c) whose effect on the production cross-section is not kinematically suppressed relative to WWF , the relative importance of its contribution is larger than that of ${\cal O}_{V\nu,AB}$.

\vskip 0.1in

\noindent{${\cal O}_{W,AB}$ and ${\cal O}_{B,AB}$}

\vskip 0.1in

The neutrino dipole operators ${\cal O}_{W,AB}$ and ${\cal O}_{B,AB}$ contribute to Higgs production via diagram \ref{fig:nur}(a) wherein the exchanged gauge boson is either a $Z^0$ or a $\gamma$ and the final state contains a neutrino and an antineutrino that are either both right-handed or both  left-handed.  The insertion of ${\cal O}_{W,AB}$ in diagrams \ref{fig:nur}(b) and (c) only contain the $W^\pm$ boson; they contribute to the same final states does ${\cal O}_{\tilde{V},AB}$ .  Note that since ${\cal O}_{B,AB}$  contributes only through  \ref{fig:nur}(a), its contribution will be suppressed relative to that of ${\cal O}_{W,AB}$.  Again, this feature can be seen from Fig.~\ref{fig:nur500Gev}.

\subsection{Flavor Nonconserving Operators}
\label{sec:flavc}

Now, we consider the case $A\ne B$ for those operators having the potentially largest effects in the flavor conserving channels: ${\cal O}_{VR,AB}$,  ${\cal O}_{VL,AB}$, and  ${\cal O}_{VL\tau,AB}$.  Here, we have two distinct cases, $A$ or $B=e$, and both $A$, $B\ne e$.  The latter case can only contribute through diagram \ref{fig:nonur}(b), whose effect is kinematically suppressed.   Hence, we ignore this case. For all three of these flavor nonconserving operators, Higgs production can occur through diagrams \ref{fig:nonur}(b), and (c) or (d), giving a final state containing $e^{\pm} \mu^{\mp}$ or $e^{\pm} \tau^{\mp}$.  Although diagrams \ref{fig:nonur}(b) (in the case of ${\cal O}_{VL,AB}$ or  ${\cal O}_{VL\tau,AB}$) and (c), and (d) (for ${\cal O}_{VL\tau,AB}$ only) could also contribute to the missing energy final state, given the small number of events involved (to be seen in Section \ref{sec:oplimits} ), we consider only the final states with charged leptons, due to their unique flavor-nonconserving signature. Results for the case $C v^2 / \Lambda^2 =10^{-2}$ are shown in Table \ref{table:fcncs} in units of $\mbox{ab}^{-1}$.  For a linear collider with $1 \, \mbox{ab}^{-1}$ of data, these numbers can be interpreted as numbers of events.

\begin{table}
\caption{Cross-sections for flavor-nonconserving processes $e^+ e^- \rightarrow H e^{\pm} l^{\mp}$, $l=\mu,\tau$ for $Cv^2/\Lambda^2=10^{-2}$.  Both charge combinations are included.  Results are in units of $10^{-6}$ pb.}
\label{table:fcncs}
\begin{tabular}{cccc|ccc }
& & $\sqrt{s}=500$ GeV & & & $\sqrt{s}=1$ TeV & \\
$m_H$ & $100$ GeV & $250$ GeV  & $400$ GeV & $100$ GeV & $300$ GeV & $500$ GeV\\
\hline
${\cal O}_{VR,e\ell}$ & $3.4$ & $0.72$ & $0.024$ & $28.$ & $14.$ & $4.2$\\  
${\cal O}_{VL,e\ell}$, ${\cal O}_{VL\tau,e\ell}$ & $3.2$ & $0.67$ & $0.023$ & $27.$ & $13.$ & $4.1$\\
\end{tabular}
\end{table}

\section{Limits on Operator Coefficients}
\label{sec:oplimits}

Precision electroweak data constrains the magnitude of many of the $C_6^j v^2/\Lambda^2$ to be  considerably smaller than the $10^{-2}$ reference value used in Section \ref{sec:newhiggs}. Constraints on  a subset of the Class B operator coefficients have been obtained using data from LEP $Z^0$-pole data\cite{Barbieri:1999tm} and from a wider array of precision electroweak observables that includes studies at LEP2 and low-energy experiments\cite{Han:2004az}. Both analyses relied on the assumption of U(3)$^5$ symmetry and \cite{Han:2004az} performed fits to EWPO including the effects of more than one operator simultaneously.

Here, we up-date these earlier analyses in a way that focuses on the Class B and Class C operators with the potentially largest effects in Higgs production. For the Class B case, these operators are ${\cal O}_{VR,\, ee}$, ${\cal O}_{VL,\, ee}$, and ${\cal O}_{VL\tau,\, ee}$. For the Class C operators, the direct experimental limits on the coefficient of ${\cal O}_{{\tilde V},\, AB}$ are weaker than our reference value of $10^{-2}$. Since the effect of this operator is quadratic in the corresponding coefficient, any significant increase in its value could lead to a several percent effect in the missing energy channel. We discuss the direct experimental and indirect constraints on these operators below. 

In order to obtain constraints on ${\cal O}_{VR,\, ee}$, ${\cal O}_{VL,\, ee}$, and ${\cal O}_{VL\tau,\, ee}$, we have performed a fit to EWPO using the GAPP routine\cite{Erler:1999ug}. The precision observables included in this fit include the data collected from $Z^0$ pole studies at LEP and SLD and a variety of low-energy precision observables, including cesium atomic parity violation\cite{Wood:1997zq}, parity-violating M\o ller scattering\cite{Anthony:2003ub}, elastic neutrino-electron scattering\cite{Vilain:1994qy} and deep inelastic neutrino-nucleus scattering\cite{Zeller:2001hh}  (for a complete list of EWPO used, see Ref.~\cite{Yao:2006px}). We have used the value $171.4 \pm 2.1$ GeV given in \cite{Brubaker:2006xn} for $M_t$.

For each operator, we derive bounds on the corresponding $C_6^j v^2/\Lambda^2$ by including both the direct contributions to a given observable as well as indirect effects that enter through modifications of the SM input parameters. The  ${\cal O}_{VL\tau,\, ee}$, for example, contains both neutral and charged current components. The neutral current component modifies the coupling of LH electrons to the $Z^0$ and enters all $e^+ e^-$ annihilation observables as well as those involving low energy parity violating processes. The charge current component contributes to the amplitude for muon decay. Inclusion of the latter contribution modifies the value of the Fermi constant, $G_\mu$, extracted from the experimental muon lifetime and that is used to normalize all electroweak amplitudes in the SM. It also indirectly affects the value of $\sin^2{\hat\theta}_W(M_Z)$ that is a derived quantity in the SM given $G_\mu$, $\alpha$, and $M_Z$ as inputs. 

Our procedure differs that followed by Refs.~\cite{Barbieri:1999tm,Han:2004az} in a few respects. First, we do not assume a U(3)$^5$ symmetry that relates operators involving different fermion generations. For example,  ${\cal O}_{VR,\, ee}$ and ${\cal O}_{VR,\, \mu\mu}$ are treated as distinct. Although it is quite reasonable to assume that flavor-dependent effects from physics above the scale $\Lambda$ are determined by Yukawa interactions (as in models with minimal flavor violation) and are, thus, suppressed, we will not make that assumption here. Second, the fits performed in Refs.~\cite{Barbieri:1999tm,Han:2004az} allowed for the simultaneous contribution from multiple effective operators and were correspondingly performed for a fixed value of $m_H$. Here, we instead include the effect of only one operator and allow the value of $m_H$ to remain a fit parameter. 


The results for the three most important Class B operators are given in Table \ref{table:gapplimits}, where we show the $1\sigma$ results and 95\% C.L. ranges for the $C_6^j v^2/\Lambda^2$ in the second and third columns, respectively.  In the last column, we give the fit results for $m_H$; for comparison, an SM fit, with the $C_6^j$ set to $0$, gives $m_H = 84 +33 -24$ GeV.  We find that inclusion of the operator containing $e_R$ fields tends to lower the best fit value for $m_H$, although it still falls within $2\sigma$ of the direct search lower bound, $m_H=114.4$ GeV. In contrast, the two operators containing first generation lepton doublet fields increases the best fit value for $m_H$.
\begin{table}
\caption{Bounds on coefficents $C_6^j$ of the $n=6$ leptonic operators obtained implied by electroweak precision observables (EWPO). First column lists the operator. Second column gives result for $C_6^j v^2/\Lambda^2$ obtained from  fit to all EWPO using the GAPP routine\cite{Erler:1999ug}. Third column gives the 95\% C.L. range on $C_6^j v^2/\Lambda^2$, while the last column gives the corresponding fit values for the Higgs mass, $m_H$. }
\label{table:gapplimits}
\begin{tabular}{c|c|c|c}
Operator & $C_6^j v^2/\Lambda^2$ & 95\% C.L. range & $m_H$  \\ 
\hline
$O_{VR,ee}$ & $-0.00037 \pm 0.00041$ &$-0.0012, \rightarrow 0.00044 $& $72 + 35 -24$ GeV  \\
$O_{VL,ee}$ & $0.00053\pm 0.00035$ &$-0.00015  \rightarrow 0.0012$ & $95 + 38 -28$ GeV \\
$O_{VL\tau,ee}$ & $0.00039 \pm 0.00039$ & $-0.00036  \rightarrow 0.0011  $ & $90 +36 -26  $ GeV   \\
\end{tabular}
\end{table}
We also observe that the constraints given in Table \ref{table:gapplimits} are somewhat weaker than those obtained in Ref.~\cite{Han:2004az}, presumably because we have not invoked a U(3)$^5$ symmetry and have allowed the value of $m_H$ to vary\footnote{In the notation of Ref.~\cite{Han:2004az}, the operators ${\cal O}_{VR,\, ee}$, ${\cal O}_{VL,\, ee}$, and ${\cal O}_{VL\tau,\, ee}$ correspond to ${\cal O}_{he}$, ${\cal O}_{h\ell}^s$, and ${\cal O}_{h\ell}^t$ when a U(3)$^5$ symmetry is assumed.}. The results of our fit -- together with the analysis of Section \ref{sec:newhiggs} -- thus, indicate the largest possible effects that one might anticipate for Class B operators. 

We have also checked that EWPO do not allow the $|C_6^j v^2/\Lambda^2|$ to be large than $10^{-2}$ for the other flavor-conserving Class B operators by considering the $Z^0$ pole observables alone and comparing SM predictions for a range of $m_H$ with the results obtained from LEP and SLD. To this end, we obtain the SM predictions using ZFITTER \cite{Bardin:1999yd} \cite{Arbuzov:2005ma}, which  requires input values for $M_Z$, $M_t$, $m_H$, $\alpha_s(M_Z)$, and $\Delta \alpha^{(5)}_{had}$.  We take the following for our ZFITTER inputs:
\bea
\label{eq:inputs}
M_Z &=& 91.1876 \pm 0.0021 \,\mbox{GeV} \, \mbox{\cite{Yao:2006px}} \nonumber\\
M_t &=& 171.4 \pm 2.1 \,\mbox{GeV} \, \mbox{\cite{Brubaker:2006xn}} \nonumber\\
m_H &=& 200 \pm 100 \,\mbox{GeV} \\
\alpha_s(M_Z) &=& 0.1176 \pm 0.002 \,\mbox{\cite{Yao:2006px}} \nonumber\\
\Delta \alpha^{(5)}_{had} (\alpha_s(M_Z) = 0.1176) &=& 0.02772 \pm 0.0002 \nonumber
\eea
where the value for $\Delta \alpha^{(5)}_{had}$ is a linear interpolation of points given in \cite{Erler:1998sy}.  The range on $m_H$ is chosen to be (possibly artificially) large to accomodate any possibility that the current upper bounds on $m_H$ could be evaded with the addition of the operators ${\cal O}_{6,j}$.  The authors of \cite{Barbieri:1999tm} find, for a particular Higgs mass, ranges of the operator coefficients  for which $\chi^2-\chi^2_{min}<3.85$, where $\chi^2_{min}$ is the $\chi^2$ of the SM fit with the operator coefficients set to zero.  They find values of the coefficients of ${\cal O}_{VR}$ and ${\cal O}_{VL\tau}$ which satisfy this criterion for values of $m_H$ as high as $300$ GeV.  Even when we include the error for this broad range of Higgs mass, we still find limits on the operator coefficients that are tighter than our reference value of $10^{-2}$.

These yield the following predictions for the SM observables:
\bea
\Gamma(Z\rightarrow \mbox{inv}) &=& 501.399 +0.216 -0.201 \,\mbox{MeV} \nonumber \\
\Gamma(Z\rightarrow e^+ e^-) &=& 83.932 +0.053 -0.044\, \mbox{MeV} \nonumber \\
\Gamma(Z\rightarrow \mu^+ \mu^-) &=& 83.932 +0.053 -0.044\, \mbox{MeV} \nonumber \\
\Gamma(Z\rightarrow \tau^+ \tau^-) &=&  83.742 +0.053 -0.044\,\mbox{MeV}. \nonumber 
\eea
The errors on these values were obtained by separately computing the errors due to the uncertainties on the input parameters given in Eqs. (\ref{eq:inputs}) and adding them in quadrature.  The asymmetry in the errors is due to the dependence of the results on  $\ln{m_H}$.

These predictions are to be compared with the experimental values for the $Z$ widths and branching fractions \cite{Yao:2006px}:
\bea
\Gamma(Z\rightarrow \mbox{inv}) &=& 499.0 \pm  1.5 \,\mbox{MeV} \nonumber \\
\Gamma(Z\rightarrow e^+ e^-) &=& 83.91 \pm 0.12  \, \mbox{MeV} \nonumber \\
\Gamma(Z\rightarrow \mu^+ \mu^-) &=& 83.99 \pm 0.18 \, \mbox{MeV} \nonumber \\
\Gamma(Z\rightarrow \tau^+ \tau^-) &=&  84.08 \pm 0.22 \, \mbox{MeV} \nonumber \\
BR(Z\rightarrow e^{\pm} \mu^{\mp}) &=& < 1.7 \times 10^{-6} \, \mbox{at} \, 95\% \, \mbox{CL}\nonumber \\
BR(Z\rightarrow e^{\pm} \tau^{\mp}) &=& <  9.8 \times 10^{-6} \, \mbox{at} \, 95\% \, \mbox{CL}\nonumber \nonumber
\eea

The largest source of theoretical error in the SM predictions, as well as the asymmetry in the theoretical error, arises from the range taken for $m_H$.  However, the experimental error dominates over the theoretical error for all of the above observables. The resulting bounds on the $Cv^2/\Lambda^2$ for the Class B operators are given in Table \ref{table:oplimits}. We do not include bounds on the ${\cal O}_{VR,ee}$, ${\cal O}_{VL,ee}$, and ${\cal O}_{VL\tau,ee}$  operators in this table because the GAPP fit provides significantly tighter limits than using the $Z$ partial widths alone.


From the limits on the branching fractions of the Z to $e^{\pm} \mu^{\mp}$ and  $e^{\pm} \tau^{\mp}$, we can deduce limits on the coefficients for ${\cal O}_{VR,AB}$,  ${\cal O}_{VL,AB}$, and  ${\cal O}_{VL\tau,AB}$, where $A\ne B$ and $A$ or $B=e$.  We obtain
\begin{eqnarray}
\left| \frac{C_{e \mu} v^2}{\Lambda^2} \right| &<& 0.0071 \nonumber \\
\left| \frac{C_{e \tau} v^2}{\Lambda^2} \right| &<& 0.017 \\
\end{eqnarray}
at $95\%$ CL for all three operators.  As these coefficients enter into the cross-sections for these processes quadratically, we can see from Table \ref{table:fcncs} that these limits allow, for example, as many as $\sim 80$ $H e^{\pm} \tau^{\mp}$ events for a Higgs in the low-mass region at a linear collider with $\sqrt{s}=1$ TeV.  It will be interesting to explore the feasibility of observing these events at a Linear Collider.

\begin{table}
\caption{$95\%$ CL intervals on the coefficents $C_6^j$ of the 6D leptonic operators, multiplied by $v^2/{\Lambda^2}$.  In the case of ${\cal O}_{\nu_R,AB}$, the limit is instead on $\sum_{A,B} \left|C_{\nu_R}^{AB}\right|^2 v^4/\Lambda^4$.}
\label{table:oplimits}
\begin{tabular}{l|ll }
Operator & $Min(\frac{C^j v^2}{\Lambda^2})$ & $Max(\frac{C^j v^2}{\Lambda^2})$  \\
\hline
${\cal O}_{VR,\mu\mu} $  & $-0.0027 $& $0.0020$ \\
${\cal O}_{VR,\tau\tau} $  & $-0.0050$& $0.0007$\\
${\cal O}_{VR,e\mu}$ & $-0.0071$ & $0.0071$ \\
${\cal O}_{VR,e\tau}$ & $-0.017$ & $0.017$\\
${\cal O}_{VL,\mu\mu} $  & $-0.0017$ & $0.0023$ \\ 
${\cal O}_{VL,\tau\tau} $  & $-0.0006$ & $0.0043$ \\ 
${\cal O}_{VL,e\mu}$ & $-0.0071$ & $0.0071$ \\
${\cal O}_{VL,e\tau}$ & $-0.017$ & $0.017$\\
${\cal O}_{VL\tau,\mu\mu} $ & $-0.0039$ & $0.0054$ \\
${\cal O}_{VL\tau,\tau\tau} $ & $-0.0006$ & $0.0043$ \\
${\cal O}_{VL\tau,e\mu}$ & $-0.0071$ & $0.0071$ \\
${\cal O}_{VL\tau,e\tau}$ & $-0.017$ & $0.017$\\
${\cal O}_{\nu_R,AB}$ & & $<.0068$  \\

\end{tabular}
\end{table}

Some, but not all, of the Class C operators are also constrained by EWPO. To constrain $C_{V\nu,AB}$, we consider the contribution of ${\cal O}_{V\nu,AB}$ to the invisible width of the $Z$ boson, $\Gamma_{\rm inv}$.  Although the measured value of $\Gamma_{\rm inv}$ disagrees slightly with the SM prediction (the experimental value is $1.6\sigma$ below the SM expectation) , ${\cal O}_{V\nu,AB}$ cannot explain this small discrepancy, as it does not interfere with the SM process and can only increase the cross-section for $Z\rightarrow \nu \bar{\nu}$.  We calculate the limit on this operator using the procedure for obtaining one-sided confidence level intervals given in Ref.~\cite{Feldman:1997qc}. \\

For the remaining operators, all of which contain $\nu_R$, we consider first direct experimental constraints. For example, the operator ${\cal O}_{\tilde{V},eB}$ also contributes to the Michel spectrum for the decay of polarized muons. From the recent global analysis of  muon decay measurements reported in  Ref.~\cite{Gagliardi:2005fg} we obtain
\be
\left\vert C_{{\tilde V},\, eB} v^2 / \Lambda^2\right| \leq 0.208
\ee
at 90 \% C.L. In contrast to the situation with the Class B operators and ${\cal O}_{V\nu,AB}$, the direct constraints on ${\cal O}_{\tilde{V},eB}$ are considerably weaker than our benchmark $10^{-2}$ value for $C_6^j v^2/\Lambda^2$. Considerably more stringent expectations can be obtained by observing that ${\cal O}_{\tilde{V},eB}$ contributes to the $n=6$ neutrino mass operator ${\cal O}^\nu_{M,\, AB}$ through radiative corrections. A complete renormalization group analysis of the mixing between these operators was carried out in Ref.~\cite{Erwin:2006uc}. In order to avoid \lq\lq unnatural" fine tuning, the radiative contributions to the neutrino mass matrix element $m_\nu^{AB}$ due to ${\cal O}_{V\nu,AB}$ cannot be substantially larger than the scale of neutrino mass itself. Using an upper bound of 1 eV for this scale we obtain the following naturalness bound on ${C_{\tilde{V},eB} v^2}/{\Lambda^2}$
\be
\left| \frac{C_{\tilde{V},eB} v^2}{\Lambda^2} \ln \frac{v}{\Lambda} \right|  <  (0.5-3) \times 10^{-3}.
\label{eq:mnubounds}
\ee
where the range on $C_{\tilde{V},eB}$ corresponds to $114 \, \mbox{GeV} < m_H < 185 \, \mbox{GeV}$. The latter affects the renormalization group analysis since the entries in the anomalous dimension matrix depend on the Higgs boson quartic self coupling, $\lambda=m_H^2/2v^2$. 

The coefficients of the magnetic moment operators are bounded by upper limits on
neutrino magnetic moments that range from $10^{-10}$ to $10^{-12}$ Bohr
magnetons \cite{Raffelt:1999gv,Sutherland:1975dr,Xin:2005ky,Daraktchieva:2005kn,Liu:2004ny,Beacom:1999wx}.  Taking the upper limit of
these bounds implies that $|C_{W,AB} v^2/\Lambda^2|$ and $|C_{B,AB}
v^2/\Lambda^2|$ are no larger than $\sim 10^{-5}$. Neutrino mass naturalness 
considerations imply bounds that are roughly four orders of magnitude
more stringent than those obtained directly from magnetic moment limits. Either way, the effects of these operators on Higgs production
will be unobservable.

\section{Discussion and Conclusions}
\label{sec:conclusions}

The bounds we obtain on the operator coefficients generally satisfy  $|Cv^2/\Lambda^2|<10^{-2}$, implying smaller corrections to the Higgs production cross-sections than those given in Figures \ref{fig:2lree500}-\ref{fig:nur500Gev}, for which we have used $Cv^2/\Lambda^2 = 10^{-2}$.  Nevertheless, comparing the bounds on $|Cv^2/\Lambda^2|$ for ${\cal O}_{VR,ee}$, ${\cal O}_{VL,ee}$, and ${\cal O}_{VL\tau,ee}$ with the results in Figures \ref{fig:2lree500}, \ref{fig:2Lee01}, and \ref{fig:2Ltauee01}, we see that the interference with the SM HZ process can be substantial in the $Hf{\bar f}$ channel with $f=\mu$, $\tau$, or $q$, with corrections of more than 5\% (20\%) allowed for $\sqrt{s}=500$ GeV (1 TeV).  The relative impact of these operators on the $He^+e^-$ and $H+{\not\!\! E}$ channels is considerably smaller, since the SM cross-section receives large WWF and ZZF contributions.    Additionally, we have checked the non-interference contributions of these operators and find that, for $|C v^2/\Lambda^2|=10^{-3}$ (toward the upper end of the $95\%$ CL range) the non-interference terms can contribute an additional $3\%$ to the $Hf{\bar f}$ cross-section for $\sqrt{s}=1$ TeV.  The contributions of the non-interference terms to the $Hf{\bar f}$ channel at $\sqrt{s}=500$ GeV and to the $H+{\not\!\! E}$ and $He^+ e^-$ channels at either $\sqrt{s}$ are all $< 1\%$. 

Conversely, despite the less stringent limits on their coefficients, the operators ${\cal O}_{VR, AA}$, ${\cal O}_{VL,AA}$, and ${\cal O}_{VL\tau,AA}$ for $A=\mu$, $\tau$, or $q$ cannot generate significant corrections to the $HA{\bar A}$ production cross-section, due to the kinematic suppression of the corresponding interference amplitude relative to SM HZ.

In the case of the Class C operators, which contribute only to the  $H+{\not\!\! E}$ channel, the magnitude of possible corrections is generally  smaller than $10^{-3}$ of the SM cross-section, assuming $Cv^2/\Lambda^2=10^{-2}$. Amplitudes containing these operators do not interfere with SM amplitudes as they contain RH neutrino states, so the quadratic dependence of their contribution to the cross-section on the operator coefficients can lead to considerable suppression. From our analysis of the limits in Section \ref{sec:oplimits}, we conclude that for ${\cal O}_{V\nu_R,AB}$, whose coefficient is constrained by the invisible width of the $Z^0$, the possible effect is negligible. A similar conclusion applies to ${\cal O}_{W}$ and ${\cal O}_{B}$, which are constrained by limits on neutrino magnetic moments.  For the operator ${\cal O}_{\tilde V,eB}$, the constraint on the coefficient implied from the $\mu$-decay Michel spectrum is more than an order of magnitude weaker than assumed in obtaining Figure \ref{fig:nur500Gev}, and would allow the corresponding correction to the missing energy channel to be of order 10\% or more (recall that the dependence on the coefficient is quadratic). On the other hand, the bound obtained from neutrino mass naturalness considerations is substantially smaller than $|Cv^2/\Lambda^2|=10^{-2}$, suggesting an unobservable contribution from this operator to the $H+{\not\!\! E}$ cross-section. Thus, the observation of a deviation in this channel without similar deviations in the $Hq\bar{q}$ and $H\ell\bar{\ell}$ channels-- though unlikely -- would imply the presence of fine tuning in order to avoid unacceptably large radiative contributions to neutrino mass.

Summarizing the situation more broadly, we find that there exists considerably less room for effects on Higgs production from higher dimension operators containing fermions than from purely bosonic operators. Constraints from EWPO generally imply $|C v^2/\Lambda^2| << 10^{-2}$. The impact of this suppression can be overcome only in channels that are dominated by SM HZ due to the absence of an off-shell $Z^0$-boson propagator in amplitudes containing any of the operators ${\cal O}_{VR,ee}$, ${\cal O}_{VL,ee}$, and ${\cal O}_{VL\tau,ee}$. In contrast, purely bosonic operators, such as $\partial^\mu(\phi^\dag\phi)\partial_\mu(\phi^\dag\phi)$, can lead to potentially significant deviations in a variety of channels simultaneously, since (a) they affect the couplings of the Higgs to gauge bosons and (b) the constraints from EWPO are weak\cite{Barger:2003rs}.  A comprehensive study of Higgs production in a variety of channels at a linear collider would allow one to disentangle possible effects from different classes of effective operators, thereby providing new clues about physics at high scales\footnote{Studies of polarization observables or angular distributions may also allow one to distinguish the effects of different effective operators, along the lines suggested in Ref. \cite{Barger:1993wt}. We thank V. Barger for bringing this possibility to our attention.}.

\acknowledgments
The authors are particularly indebted to J. Erler for making the GAPP code available and for several helpful discussions about fits to electroweak precision observables. We also
 thank V. Barger, N. Bell, V. Cirigliano, M. Gorshteyn, T. Han, P. Langacker,  P. Vogel, and M. Wise for helpful discussions. This work was supported in part under U.S. Department of Energy contracts FG02-05ER41361 and DE-FG03-ER40701 and National Science Foundation award PHY-0555674.

\bibliographystyle{h-physrev}


\begin{thebibliography}{99}


\bibitem{Heinemeyer:2005gs}
  S.~Heinemeyer {\it et al.},
  arXiv:hep-ph/0511332.

\bibitem{Barger:2003rs}
  V.~Barger, T.~Han, P.~Langacker, B.~McElrath and P.~Zerwas,
  Phys.\ Rev.\ D {\bf 67}, 115001 (2003)
  [arXiv:hep-ph/0301097].

\bibitem{Manohar:2006gz}
  A.~V.~Manohar and M.~B.~Wise,
  Phys.\ Lett.\ B {\bf 636}, 107 (2006)
  [arXiv:hep-ph/0601212].

\bibitem{Grinstein:2007iv}
  B.~Grinstein and M.~Trott,
  arXiv:0704.1505 [hep-ph].




\bibitem{Black:2002wh}
  D.~Black, T.~Han, H.~J.~He and M.~Sher,
  Phys.\ Rev.\  D {\bf 66}, 053002 (2002)
  [arXiv:hep-ph/0206056].



\bibitem{Barbieri:1999tm}
  R.~Barbieri and A.~Strumia,
  Phys.\ Lett.\ B {\bf 462}, 144 (1999)
  [arXiv:hep-ph/9905281].

\bibitem{Han:2004az}
  Z.~Han and W.~Skiba,
  Phys.\ Rev.\ D {\bf 71}, 075009 (2005)
  [arXiv:hep-ph/0412166].


\bibitem{Bell:2005kz}
  N.~F.~Bell, V.~Cirigliano, M.~J.~Ramsey-Musolf, P.~Vogel and M.~B.~Wise,
  Phys.\ Rev.\ Lett.\  {\bf 95}, 151802 (2005)
  [arXiv:hep-ph/0504134].

\bibitem{Erwin:2006uc}
  R.~J.~Erwin, J.~Kile, M.~J.~Ramsey-Musolf and P.~Wang,
  arXiv:hep-ph/0602240.

Moortgat-Pick et al: hep-ph/0507011
\bibitem{Moortgat-Pick:2005cw}
  G.~A.~Moortgat-Pick {\it et al.},
  arXiv:hep-ph/0507011.

\bibitem{Han:1999xd}
  T.~Han, T.~Huang, Z.~H.~Lin, J.~X.~Wang and X.~Zhang,
  Phys.\ Rev.\  D {\bf 61}, 015006 (2000)
  [arXiv:hep-ph/9908236].



\bibitem{Gunion:1989we}
  J.~F.~Gunion, H.~E.~Haber, G.~L.~Kane and S.~Dawson,
{\it The Higgs Hunter's Guide}, (Westview Press, 2001).


\bibitem{Desch:2001at}
  K.~Desch and N.~Meyer,
LC-PHSM-2001-025
{\it  In *2nd ECFA/DESY Study 1998-2001* 1694-1704}

\bibitem{Garcia-Abia:1999kv}
  P.~Garcia-Abia and W.~Lohmann,
  Eur.\ Phys.\ J.\ directC {\bf 2}, 2 (2000)
  [arXiv:hep-ex/9908065].

\bibitem{Garcia-Abia:2005mt}
  P.~Garcia-Abia, W.~Lohmann and A.~Raspereza,
  arXiv:hep-ex/0505096.

\bibitem{Meyer:2004ha}
  N.~Meyer and K.~Desch,
  Eur.\ Phys.\ J.\ C {\bf 35}, 171 (2004).

\bibitem{Leung:1984ni}
  C.~N.~Leung, S.~T.~Love and S.~Rao,
  Z.\ Phys.\  C {\bf 31}, 433 (1986).



\bibitem{Buchmuller:1985jz}
  W.~Buchmuller and D.~Wyler,
  Nucl.\ Phys.\ B {\bf 268}, 621 (1986).



\bibitem{Pukhov:1999gg}
  A.~Pukhov {\it et al.},
  arXiv:hep-ph/9908288.

\bibitem{Pukhov:2004ca}
  A.~Pukhov,
  arXiv:hep-ph/0412191.

\bibitem{Yao:2006px}
  W.~M.~Yao {\it et al.}  [Particle Data Group],
  J.\ Phys.\ G {\bf 33}, 1 (2006).


\bibitem{Erler:1999ug}
 J.~Erler,
 arXiv:hep-ph/0005084.

\bibitem{Wood:1997zq}  
C.~S.~Wood, S.~C.~Bennett, D.~Cho, B.~P.~Masterson, J.~L.~Roberts, C.~E.~Tanner and C.~E.~Wieman, 
  Science {\bf 275}, 1759 (1997).

\bibitem{Anthony:2003ub}
  P.~L.~Anthony {\it et al.}  [SLAC E158 Collaboration],
  Phys.\ Rev.\ Lett.\  {\bf 92}, 181602 (2004)
  [arXiv:hep-ex/0312035].

\bibitem{Vilain:1994qy}
  P.~Vilain {\it et al.}  [CHARM-II Collaboration],
  Phys.\ Lett.\  B {\bf 335}, 246 (1994).


\bibitem{Zeller:2001hh}
  G.~P.~Zeller {\it et al.}  [NuTeV Collaboration],
  Phys.\ Rev.\ Lett.\  {\bf 88}, 091802 (2002)
  [Erratum-ibid.\  {\bf 90}, 239902 (2003)]
  [arXiv:hep-ex/0110059].




\bibitem{Brubaker:2006xn}
  E.~Brubaker {\it et al.}  [Tevatron Electroweak Working Group],
  arXiv:hep-ex/0608032.

\bibitem{Bardin:1999yd}
  D.~Y.~Bardin, P.~Christova, M.~Jack, L.~Kalinovskaya, A.~Olchevski, S.~Riemann and T.~Riemann,
  Comput.\ Phys.\ Commun.\  {\bf 133}, 229 (2001)
  [arXiv:hep-ph/9908433].



\bibitem{Arbuzov:2005ma}
  A.~B.~Arbuzov {\it et al.},
  Comput.\ Phys.\ Commun.\  {\bf 174}, 728 (2006)
  [arXiv:hep-ph/0507146].



\bibitem{Erler:1998sy}
  J.~Erler,
  Phys.\ Rev.\  D {\bf 59}, 054008 (1999)
  [arXiv:hep-ph/9803453].



\bibitem{Feldman:1997qc}
  G.~J.~Feldman and R.~D.~Cousins,
  Phys.\ Rev.\ D {\bf 57}, 3873 (1998)
  [arXiv:physics/9711021].

\bibitem{Gagliardi:2005fg}  C.~A.~Gagliardi, R.~E.~Tribble and N.~J.~Williams,
  Phys.\ Rev.\ D {\bf 72}, 073002 (2005)
  [arXiv:hep-ph/0509069].  


\bibitem{Beacom:1999wx}
  J.~F.~Beacom and P.~Vogel,
  Phys.\ Rev.\ Lett.\  {\bf 83}, 5222 (1999)
  [arXiv:hep-ph/9907383].

\bibitem{Liu:2004ny}
  D.~W.~Liu {\it et al.}  [Super-Kamiokande Collaboration],
  Phys.\ Rev.\ Lett.\  {\bf 93}, 021802 (2004)
  [arXiv:hep-ex/0402015].


\bibitem{Daraktchieva:2005kn}
  Z.~Daraktchieva {\it et al.}  [MUNU Collaboration],
  Phys.\ Lett.\  B {\bf 615}, 153 (2005)
  [arXiv:hep-ex/0502037].

\bibitem{Xin:2005ky}
  B.~Xin {\it et al.}  [TEXONO Collaboration],
  Phys.\ Rev.\  D {\bf 72}, 012006 (2005)
  [arXiv:hep-ex/0502001].

\bibitem{Sutherland:1975dr}
  P.~Sutherland, J.~N.~Ng, E.~Flowers, M.~Ruderman and C.~Inman,
  Phys.\ Rev.\  D {\bf 13}, 2700 (1976).

\bibitem{Raffelt:1999gv}
  G.~G.~Raffelt,
  Phys.\ Rept.\  {\bf 320}, 319 (1999).


\bibitem{Barger:1993wt}
  V.~D.~Barger, K.~m.~Cheung, A.~Djouadi, B.~A.~Kniehl and P.~M.~Zerwas,
  Phys.\ Rev.\  D {\bf 49}, 79 (1994)
  [arXiv:hep-ph/9306270].
















\end{thebibliography}

\end{document}